\documentclass[fleqn,10pt]{wlscirep}
\usepackage[utf8]{inputenc}
\usepackage[T1]{fontenc}
\usepackage{cite}
\usepackage{soul}
\usepackage{acronym}
\usepackage[caption=false,font=footnotesize]{subfig}
\usepackage{wrapfig}
\usepackage{multirow}
\usepackage{tabularx}
\newcolumntype{Y}{>{\centering\arraybackslash}X}
\newcolumntype{Z}{>{\raggedleft\arraybackslash}X}
\usepackage{booktabs}
\usepackage{threeparttable}
\usepackage[normalem]{ulem}

\usepackage{tikz}
\usetikzlibrary{shapes,arrows}
\usetikzlibrary{positioning}

\usepackage{./custom_commands}

\usepackage[crop=pdfcrop,cleanup={.tex, .dvi, .ps, .pdf, .log, .aux, .bbl, .out}]{pstool}

\title{\textcolor{blue}{Decision Support for the Quickest Detection of Critical COVID-19 Phases}}

\author[1*]{Paolo~Braca}
\author[1]{Domenico~Gaglione}
\author[2]{Stefano~Marano}
\author[1]{Leonardo~M.~Millefiori}
\author[3]{Peter~Willett}
\author[3]{Krishna~Pattipati}

\affil[1]{NATO STO Centre for Maritime Research and Experimentation, Research Department, La Spezia, 19126, Italy}
\affil[2]{University of Salerno, Department of Information \& Electrical Engineering  and Applied Mathematics (DIEM), Fisciano (SA), 84084, Italy}
\affil[3]{University of Connecticut, Department of Electrical and Computer Engineering, Storrs, 06269, USA}

\affil[*]{Corresponding author: paolo.braca@cmre.nato.int}

\renewcommand{\textcolor}[2]{#2}

\begin{abstract}
During the course of an epidemic, one of the most challenging tasks for authorities is to decide what kind of restrictive measures to introduce and when these should be enforced.
In order to take informed decisions in a fully rational manner, the onset of a critical regime, characterized by an exponential growth of the contagion, must be identified \emph{as quickly as possible}. Providing rigorous quantitative tools to detect such an onset represents an important contribution from the scientific community to proactively support the political decision makers.
In this paper, leveraging the quickest detection theory, we propose a mathematical model of the COVID-19 pandemic evolution and develop decision tools to rapidly detect the passage from a controlled regime to a critical one. A new sequential test ---referred to as MAST (mean-agnostic sequential test)---is presented, and demonstrated on publicly available COVID-19 infection data from different countries. Then, the performance of MAST is investigated for the second pandemic wave, showing an effective trade-off between average decision delay $\Delta$ and risk $R$, where $R$ is inversely proportional to the time required to declare the need to take unnecessary restrictive measures. 
\textcolor{blue}{To quantify risk, in this paper we adopt as its proxy the average occurrence rate of false alarms, in that a false alarm risks unnecessary social and economic disruption. Ideally, the decision mechanism should react as quick as possible for a given level of risk.}
We find that all the countries share the same behaviour in terms of quickest detection, specifically the risk scales exponentially with the delay, $R \sim \exp{(-\omega \Delta)}$, where $\omega$ depends on the specific nation. 
For a reasonably small risk level, say, one possibility in ten thousand (i.e., unmotivated implementation of countermeasures every 27~years, on the average), the proposed algorithm detects the onset of the critical regime with delay between a few days to three weeks, much earlier than when the exponential growth becomes evident. 
Strictly from the quickest-detection perspective adopted in this paper, it turns out that countermeasures against the second epidemic wave have not always been taken in a timely manner.
The developed tool can be used to support decisions at different geographic scales (regions, cities, local areas, etc.), levels of risk, instantiations of controlled/critical regime, and is general enough to be applied to different pandemic time-series. Additional analysis and applications of MAST are made available on a dedicated website.

\end{abstract}

\begin{document}
	
\flushbottom
\maketitle
%
%
\thispagestyle{empty}

\section*{Introduction}

With more than 57 million cases worldwide and over one and 1.36 million deaths as of November 20, 2020, the outbreak of coronavirus disease (COVID-19)\textcolor{blue}{,~\cite{COVID19_Def}} is undeniably one of the worst global crises since World War II. 
In March 2020, the exponential increase of individuals needing hospitalization in intensive care units, combined with the lack of effective cures and vaccines, pushed many governments to take extraordinary measures aimed at ``flattening the curve'' of infections.~\cite{WHO:2020_2,Anderson2020,Hellewell2020} The adopted measures included the limitation of mobility and social activities, closure of schools, universities, shops, factories, and so forth, up to the extreme act of national lockdowns. Evidence that such measures achieved a reduction of the rate of new infections are gradually appearing in the scientific literature.~\cite{MaiBenBro:K20,DehningScience2020,AdaptiveBayesian}

These measures contributed to keep the spread of COVID-19 under control for some time, but we are now, in November 2020, experiencing the onset of a new exponential growth of confirmed cases, the ``second wave,'' with severe risks for personal health and healthcare systems under severe stress. 
Governments and authorities are facing again the difficult task of deciding if and when new containment measures may be needed. In absence of limitations to mobility and social activities, the pandemic spreads exponentially in time, so that any delay in applying restrictions may lead to severe consequences. On the other hand, accounting for the social and economic impact of the possible countermeasures, already observed in the first half of 2020,~\cite{Maria2020,SHARIF2020,GlobalSupplyChain,millefiori2020covid19} restrictions should be taken only if and when it is strictly necessary. Managing the trade-off between these contrasting requirements is extremely challenging. 

To address this challenge, we leverage sequential detection theory,~\cite{poorbook,Lehmann-testing,shao} and specifically \emph{quickest detection} schemes~\cite{basseville-book,poor-book-quickest,TRUONG2020107299} to propose a rigorous methodology aimed at identifying \emph{as quickly as possible} the onset of an exponential growth of the pandemic evolution. 
The proposed procedure --- referred to as MAST (mean-agnostic sequential test) --- 
is designed to minimize the average time to detect a change in regime:~\cite{basseville-book,poor-book-quickest,TRUONG2020107299} from a situation in which the pandemic is under control (${\cal H}_0$ regime) to the onset of an exponential growth  (${\cal H}_1$ regime).

Quickest detection theory has a long history,~\cite{page} and has been successfully applied in several fields, including quality control, climate modeling, remote sensing, financial analysis, image analysis, security, signal and speech processing, and biomedical applications.~\cite{poor-book-quickest,TRUONG2020107299} In the context of public health surveillance, where the timely detection of various types of adverse health events is crucial, quickest detection techniques have found several applications,~\cite{Frisen2009} e.g., to reveal the onset and the peak of the epidemic period~\cite{Frisen2008} or that the peak is over.~\cite{Bock2008}

The approach pursued in this article is substantially different from most of the epidemiological models, based for instance on stochastic evolution of epidemic compartments~\cite{KerMckWal:J27, SkvRis:J12,HuQiaJunXiaWeiZhi:J20,MaiBenBro:K20,AdaptiveBayesian} and metapopulation networks,~\cite{Li489, Chinazzi395} where the goal is to predict the mid/long-term behavior of the outbreak.
For instance, in stochastic compartmental models, given an initial condition, the epidemic can have two outcomes: the number of infected individuals can increase, in which case we have a major outbreak, or decrease. The probability of a major outbreak can be computed,~\cite{Allen2017}
but it is of limited use in taking timely on-line decisions.

\begin{figure}

\centering
\psfragfig[width=.9\textwidth]{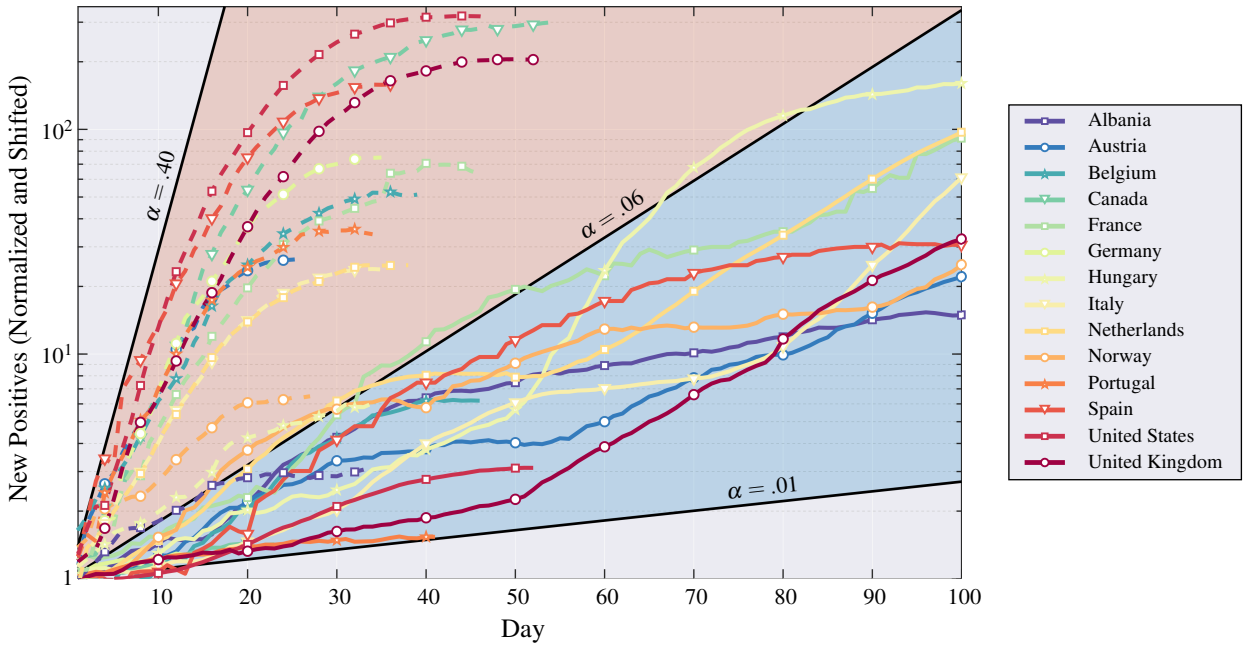}

\vspace{3mm}

\caption{Number of COVID-19 daily new positives in the first and the second wave of the pandemic for several nations. Each curve is normalized by the initial value and shifted to the same initial time. The initial point of the first wave for a specific nation is when the first positives are reported, while the initial point of the second wave is when the daily positives start to increase again --- growth rate larger than one.  
The first wave, upper region (shaded in red), was more aggressive than the second wave, mostly confined to the lower region (shaded in blue). These two regions are delimited by different values of $\alpha$ in Eq.~\eqref{eq:withalpha}. Curves of each nation are dashed if related to the first wave otherwise solid if related to the second wave.}

\label{fig:exponential-growth}

\end{figure}

The importance of taking a decision as quickly as possible in an epidemic scenario can be understood by looking at the curve of daily cases of COVID-19 infection, reported in Fig.~\ref{fig:exponential-growth} for several nations.~\cite{covid-19-JHU}
In the same figure we also report the deterministic curves of daily cases, with an exponential growth described by the following equation 
\begin{align}
    p_{n+1} = p_n(1+\alpha)  = p_1 (1+\alpha)^{n}, \quad n=1,2,\dots,
    \label{eq:withalpha}
\end{align} where $n$ is the time index (day), $1+\alpha$ is the growth rate, and $\alpha>0$, which corresponds to the ${\cal H}_1$ regime of a major outbreak. This exponential behavior is equivalent to a recently-proposed disease-transmission model,~\cite{Ashleigh2020}
and the \emph{reproduction number} defined therein is equivalent to the growth rate.
All the curves in Fig.~\ref{fig:exponential-growth} are normalized to the initial value $p_1$ and shifted to the same initial time. 
Figure~\ref{fig:exponential-growth} shows that first wave was noticeably more aggressive than the second one in terms of growth rate; specifically, in the first wave $\alpha$ was varied by country but ranged between $0.06$ and $0.40$, while in the second wave it is between $0.01$ and $0.06$. %

In the latter scenario, any delay in revealing the onset of an exponential growth phase and the consequent implementation of containment measures produces a costly exponential increase of the number of new cases.
In other words, it is essential to reduce as much as possible the decision delay to level off the curve of infection as early as possible.
At the same time, it is important to enforce restrictions only if essential, in order to avoid unnecessary social unrest and economic cost. And, inevitably, any detection procedure under the controlled regime ${\cal H}_0$ can produce false alarms, i.e., it can wrongly declare the upcoming onset of an exponential growth phase. 
The risk of taking a wrong decision can be quantified by the inverse of the mean time between false alarms. 
\textcolor{blue}{A false alarm could lead to the unnecessary adoption of restrictive measures. Since these have social and economic ramifications, adoption of this risk proxy quantifies how many times such an event occurs on average}. Balancing between detection delay and risk represents a fundamental system trade-off.
In this paper, we consider a relatively simple, but effective,  mathematical model of the pandemic and develop a decision tool to quickly detect the passage from a controlled regime to a critical one. Its effectiveness in terms of delay/risk trade-off is  demonstrated on publicly available COVID-19 data from several countries. It is our hope that the proposed MAST procedure can be useful in making timely and rational decisions to control the pandemic evolution.

\section*{Results}

\begin{figure}

\begin{minipage}{0.33\textwidth}
    \centering
    \psfragfig[width=.75\textwidth]{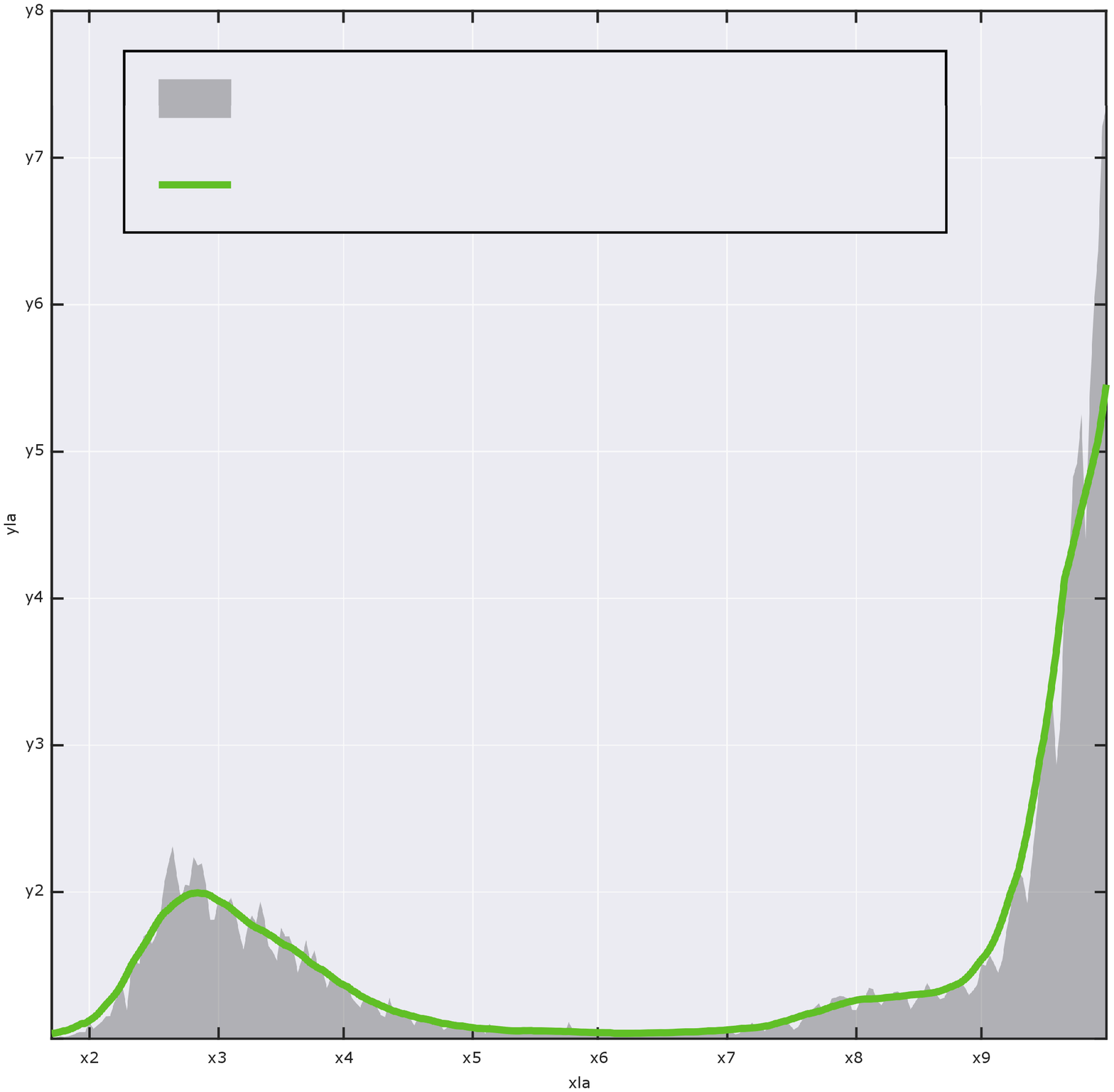} \\[4.5mm] \hspace{1.5mm} (a)
\end{minipage}%
\begin{minipage}{0.33\textwidth}
    \centering
    \psfragfig[width=0.75\textwidth]{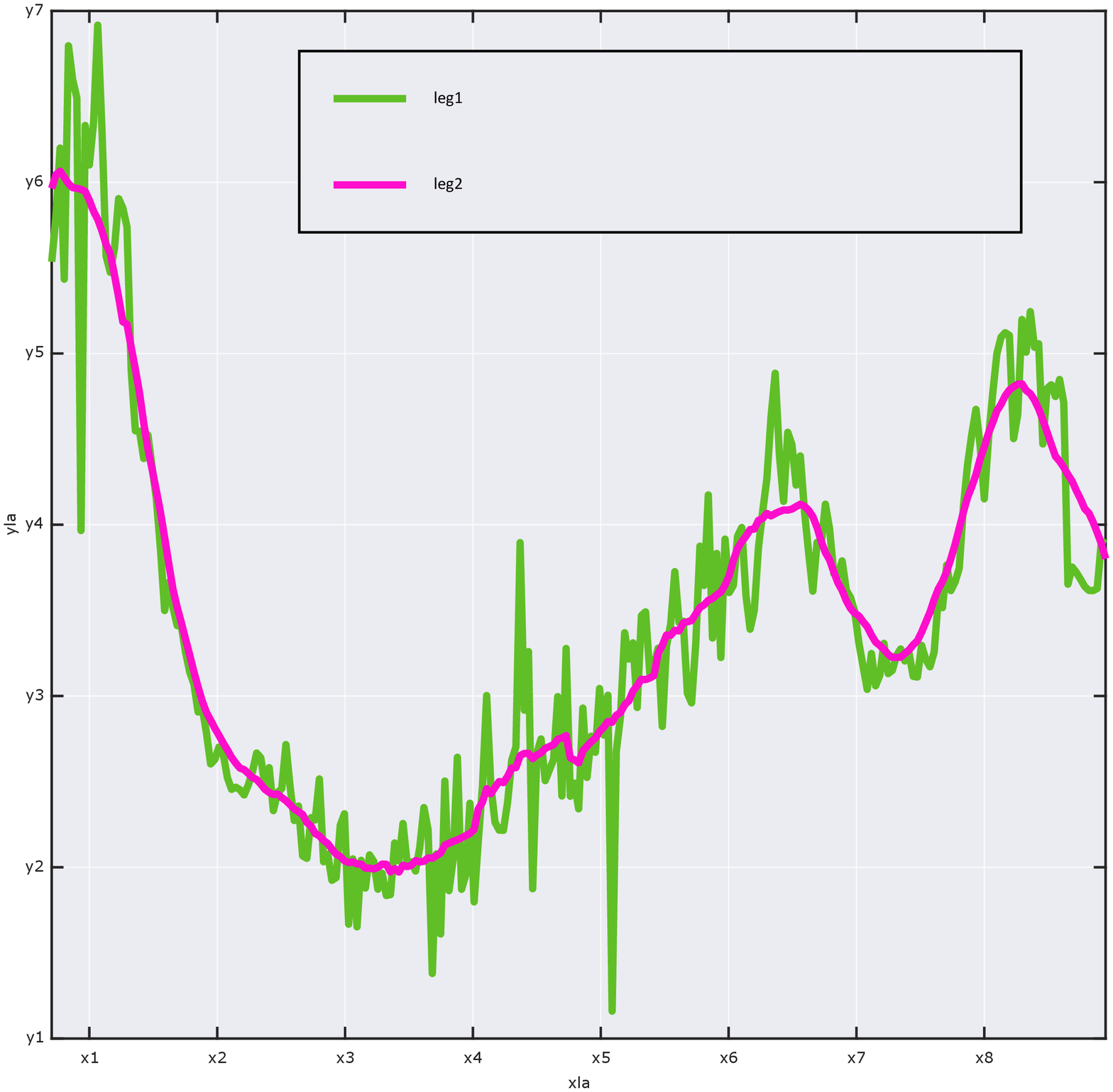} \\[4.5mm] \hspace{1.5mm} (b)
\end{minipage}%
\begin{minipage}{0.33\textwidth}
    \centering
    \psfragfig[width=0.79\textwidth]{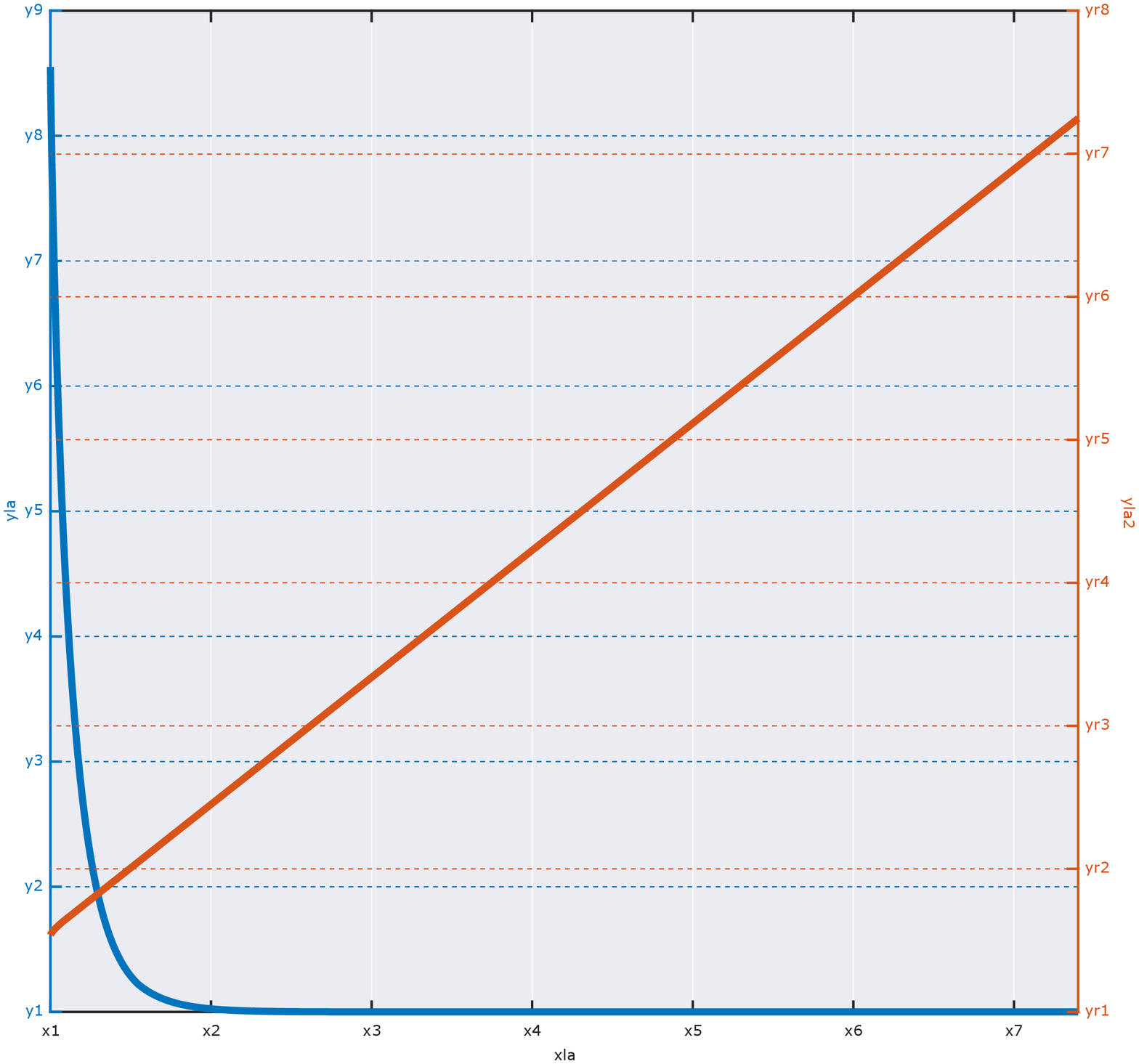} \\[4.5mm] \hspace{0mm} (c)
\end{minipage}%

\caption{(a) Daily new positive individuals in Italy since February 21, 2020, and its moving average obtained with a window of 21 days (green line).
(b) Growth rate of the epidemic computed from the averaged daily new positive cases (green line), and its time-varying mean obtained through a moving average that uses a window of 21 days (magenta line).
(c) MAST performance, in terms of risk (left axis) versus threshold and mean delay (right axis) versus threshold, obtained from the Italian data for the COVID-19 pandemic.}

\label{fig:new-positives-and-growth-rate}

\end{figure}

\begin{figure}
\centering
\psfragfig[width=0.8\textwidth]{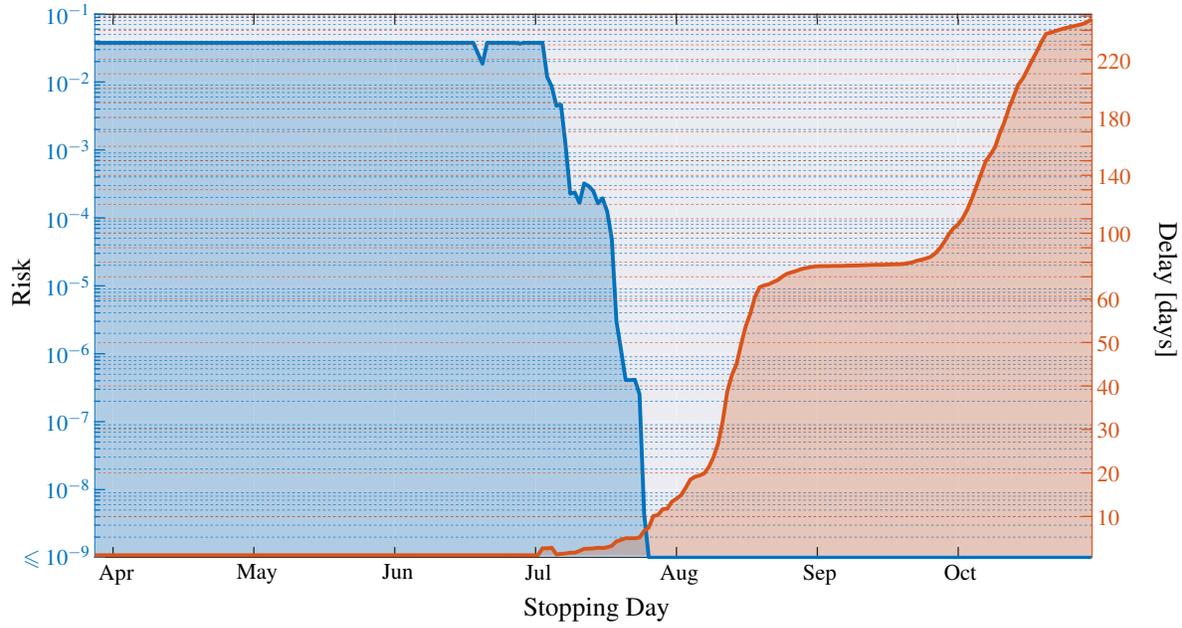}
\vspace{6mm}

\caption{Application of the MAST procedure on the COVID-19 pandemic data from Italy. On the left-side vertical axis, we select a decision threshold to correspond to a desired risk, e.g., $R=10^{-4}$. Then, the blue curve indicates the stopping day (about July 18, in the example) corresponding to the selected value of risk. Finally, the red curve referred to the right-side vertical axis shows the mean delay~$\Delta$ corresponding to the selected risk~$R$ (about 3 days).
For clarity, note that the right-side scale for the delay is split into two linear ranges, for a better rendering of the small-$\Delta$ range.}
\label{fig:test-italy}

\end{figure}

The main result of this research is the design of the MAST quickest-detection procedure, which is specifically tailored to 
epidemic scenarios, and its application on COVID-19 data. In this context, we show that the mean time $\Delta$ required to reveal 
the onset of an exponential phase is in the order of a few days (or few weeks), with a risk $R$ that scales exponentially with the delay: 
\begin{align} 
R \sim \exp(-\omega \Delta), 
\label{eq:RvsDelta}
\end{align}
where the symbol "$\sim$" refers to asymptotic  behaviour (small $R$, large $\Delta$), and $\omega$ is a parameter whose value varies from country to country. This exponential relationship holds for most \emph{optimal} quickest detection procedures under specific mathematical conditions and optimality criteria.~\cite{basseville-book} In this work, we show empirically that this optimality condition is verified by the MAST procedure when applied to the COVID-19 data, despite the fact that MAST is designed to work in the presence of uncertainty and non-stationarity of the data statistics.

To illustrate these concepts, let us refer to  Fig.~\ref{fig:new-positives-and-growth-rate}, where data from Italy are considered. Panel~(a) shows the curve of daily positives from Johns Hopkins University COVID-19 data,~\cite{covid-19-JHU} along with the smoothed version thereof, shown in green. Using the latter, we obtain the sequence of growth rates $x_1, x_2, \dots$, shown in panel~(b) (green curve) as the ratio of two consecutive values of daily positives, with its smoothed version (in magenta); see the data model in Eq.~\eqref{eq:withx}. 

In Fig.~\ref{fig:new-positives-and-growth-rate}(a), we note that the  peak of the first wave is smaller than the second one, which is likely due to the smaller number of swabs made during the first wave. In other words, in the two waves we are observing different fractions of the total infected populations. This difference does not affect our analysis because the sequence  $x_1, x_2, \dots$, of growth rates 
are ratios of successive daily positives.

As discussed in the Methods section and illustrated in the notional scheme of Fig.~\ref{fig:mast-block-diagram}, the proposed MAST procedure takes as input the sequence $x_1, x_2, \dots$ and recursively computes the sequence of decision statistics $T_1, T_2, \dots$, see Eq.~\eqref{eq:pagesquare}, where $T_n$ only depends on the observed growth rates $x_1,\dots,x_n$, up to day $n$. It is worth stressing that the growth rate process $x_1,\dots,x_n$, is assumed to be Gaussian distributed with \emph{unknown} and \emph{time-varying} mean value sequence. The MAST decision statistic is ``mean-agnostic,'' because it does not require knowledge of the time evolution of such a mean sequence, and it is therefore robust to its deterministic fluctuations. 

The onset of the critical regime is declared by the MAST at the smallest index $n$ such that  $T_n > \chi$, where $\chi$ is a threshold level. Panel~(c) of Fig.~\ref{fig:new-positives-and-growth-rate} shows the corresponding decision performance. The left axis reports the risk $R$ as a function of the threshold value $\chi$, and the right axis reports the mean delay of decision $\Delta$, again as a function of~$\chi$. 
We see that the function $R(\chi)$ is exponentially decreasing, while $\Delta(\chi)$ is linearly increasing, consistent with Eq.~\eqref{eq:RvsDelta} and with known relationships for Page's test and other quickest-detection procedures.~\cite{basseville-book,poor-book-quickest}

Combining the evidence shown in Fig.~\ref{fig:new-positives-and-growth-rate}, we obtain Fig.~\ref{fig:test-italy} that can be interpreted as follows.  
The blue curve refers to the left-side vertical axis, while the red curve refers to the right-side vertical axis.
Starting from a given level of risk selected on the left-side vertical axis, we find the \emph{stopping day} of MAST, which is the day on which the onset of the critical regime is declared (first threshold crossing). Then, referring to the red curve, one obtains the corresponding delay $\Delta$ on the right-side vertical axis. This is the mean delay incurred by the MAST procedure in declaring the onset of the critical regime. The interpretation of $\Delta$ is that the passage to the exponential growth of the pandemic takes place, on the average, $\Delta$ days before the alert produced by MAST. 
In Fig.~\ref{fig:test-italy}, values of the risk smaller than $10^{-9}$ are collapsed to $R=10^{-9}$, because for $R\le10^{-9}$ we can safely assume that risk is essentially negligible. It is worth noting that, even at negligible risk level, the stopping day is approximately July 27, and the corresponding mean delay is less than 8 days.

The analysis in Figs.~\ref{fig:new-positives-and-growth-rate}-\ref{fig:test-italy} can be repeated for other regions, yielding similar insights. We report the cases of: United States of America in Fig.~\ref{fig:test-all-cas-US}; United Kingdom in Fig.~\ref{fig:test-all-cas-UK}; France in Fig.~\ref{fig:test-all-cas-France}; and Germany in Fig.~\ref{fig:test-all-cas-Germany}. Note that in the United States, there exists also a third wave of the pandemic, which we analyze separately in Fig.~\ref{fig:test-all-cas-US}(d), by restarting the test after declaring the onset of the second wave. 
(Actually, such a third wave is likely to be a delayed second wave in different geographic regions of the U.S., as a state-by-state analysis seems to imply.)
Additional analysis on data from more geographical regions is available online~\cite{WEBSITE} and updated regularly.

\begin{figure}

\centering
\psfragfig[width=0.8\textwidth]{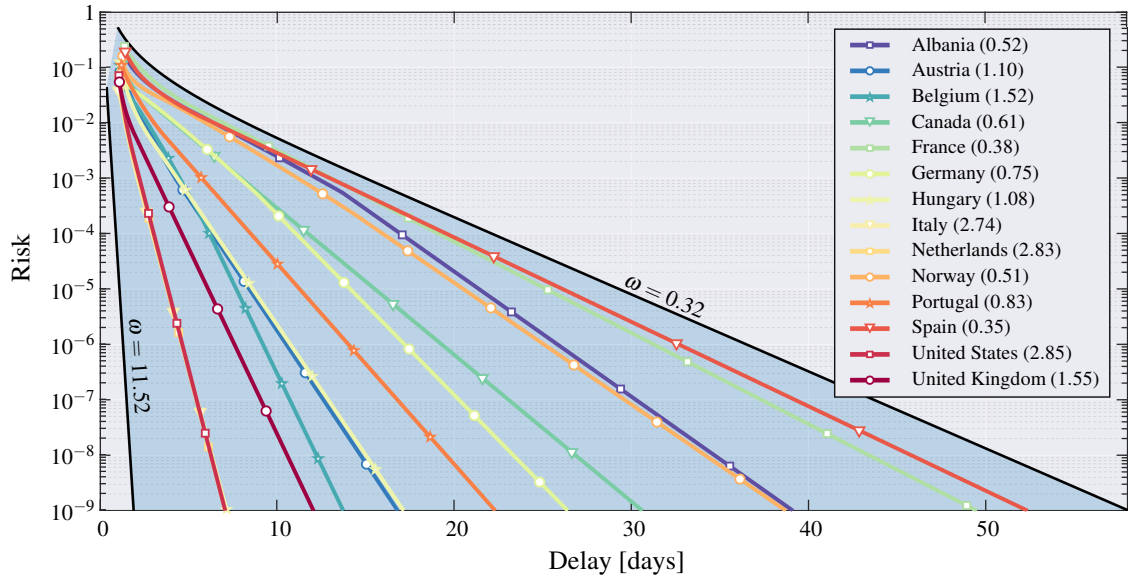}

\vspace{3mm}

\caption{Operational curve --- risk versus mean delay for decision --- for 14 Countries. For large $\Delta$, the operational curve is described by Eq.~\eqref{eq:RvsDelta}, namely $R \sim \exp(-\omega \Delta)$. 
The outermost black curves
correspond to the ideal performance of the Page's test, assuming known and constant growth rate, with $\alpha=0.01$ and $0.06$, 
respectively (extreme values of $\alpha$ for the second wave, see Fig.~\ref{fig:exponential-growth}). Each nation is characterized by a specific value of $\omega$ (reported between brackets on the legend), and all values of $\omega$ lie in the range between $\omega=0.32$ and $\omega=11.52$.}

\label{fig:operational}

\end{figure}

A comparison of the MAST performance for 14 different nations is addressed in  Fig.~\ref{fig:operational}. Different
decision performances reflect different values of the parameter $\omega$ governing the relationship between risk and delay, shown in Eq.~\eqref{eq:RvsDelta}. 
We see that, accepting a risk of $R = 10^{-4}$, the mean detection delay $\Delta$ is about 3~days for Netherlands and below 20~days for Spain. Thus, the MAST procedure detects an outbreak very expeditiously at very low risk level.

The outer black curves labeled by $\omega=0.32$ and $\omega=11.52$ represent the ideal performance of the Page's test (see Eq.~\eqref{eq:page}), in which the mean values of the growth rates $x_1, x_2,\dots$, are assumed constant, known in advance, and equal to $(1 + \alpha)$ under~${\cal H}_1$ and $(1 - \alpha)$ under~${\cal H}_0$. For such a scenario, the parameter $\omega$ appearing in  Eq.~\eqref{eq:RvsDelta} is the Kullback–Leibler distance $2\left(\alpha/\sigma\right)^2$ between the distributions under the two regimes.\cite{CT2,basseville-book} \textcolor{blue}{In signal processing language, this quantity is often referred to as signal-to-noise ratio.\cite{kaydetection}} To provide performance envelopes, we use the values $\alpha = 0.01$ and $\alpha = 0.06$ corresponding to the extreme growth rates of the second wave reported in Fig.~\ref{fig:exponential-growth}, and set $\sigma=0.025$, which is the arithmetic mean of the estimated standard deviations of the 14 countries. For the two values of $\alpha$, this yields $2\left(\alpha/\sigma\right)^2=0.32$ and $11.52$, respectively, which are the values reported in Fig.~\ref{fig:operational}.
By setting $\omega=2\left(\alpha /\sigma\right)^2$ in Eq.~\eqref{eq:RvsDelta} we observe that the higher is the growth rate, the better is the detection capability for a given risk. In other words, consistent with intuition, the more aggressive is the outbreak, the more quickly it can be detected. On the other hand, the larger is $\sigma$, the higher is the delay for a given risk. Also this effect is intuitive, because the standard deviation $\sigma$ measures the entity of random fluctuations in the data $x_1,x_2,\dots$, and reliable decisions 
require more time because the data is uncertain. These trade-offs are also observed for the MAST test run over COVID-19 data for different nations, and are captured by the parameter $\omega$ in Eq.~\eqref{eq:RvsDelta}.

\section*{Discussion}

Let us focus again on Italy, and set $R=10^{-4}$. 
Accepting a risk in the order of $R=10^{-4}$ means that there is one chance in ten thousand that the countermeasures are taken too early or, in other words,  unjustified actions against the pandemic are adopted every 27 years, on the average. By taking drastic countermeasures on July 18 --- the stopping day  prescribed by our MAST procedure for $R=10^{-4}$ --- we would have left only $\Delta \approx 3$ days of uncontrolled exponential growth of the pandemic, before addressing it.
In this respect, it should be noted that the adoption of severe countermeasures in Italy has been decided only at the beginning of November 2020. 
The delayed decision situation is analogous for other nations.


\tikzstyle{decision}=[diamond, draw, fill=gray!20, text width=4.5em, text badly centered, node distance=3cm, inner sep=0pt]
\tikzstyle{int}=[draw, fill=gray!20, minimum height=1.4cm, minimum width=2cm, align=center]
\tikzstyle{int2}=[draw, fill=white, minimum height=1.4cm, minimum width=2cm, align=center]
\tikzstyle{init} = [pin edge={to-,thin,black}]

\begin{figure}
    \centering%
    \resizebox{1.0\textwidth}{!}{%
    \begin{tikzpicture}[auto,>=latex']
        \node [draw,trapezium,trapezium left angle=70,trapezium right angle=-70,minimum height=1.4cm, minimum width=2cm, align=center, fill=gray!20] (a) {Daily\\ new cases};
        \node [int] (c) [right=1.5cm of a, pin={[init]above:$L$}] {Smoothing\\ filter};
        \node [int] (d) [right=1.5cm of c] {Growth rate\\ ${p_{n + 1}}/{p_{n}}$};
        \node [int] (f) [right=1.5cm of d] {MAST\\ statistic};
        \node [decision, pin={[init]above:$\chi$}] (h) [right=1.5 of f] {Exceeds threshold};
        \node [shape=rounded rectangle, draw, inner sep=2mm, align=center, fill=gray!20, minimum width=3cm, minimum height=1.4cm] (j) [right=1.5cm of h] {Declare\\ outbreak};
        \node [int] (k) [below=1.5cm of h] {Continue\\ observing};
        \path[->] (a) edge node {} (c);
        \path[->] (c) edge node {$p_n$} (d);
        \path[->] (d) edge node {$x_n$} (f);
        \path[->] (f) edge node {$T_n$} (h);
        \path[->] (h) edge node[above] {yes} node[below] {$\mathcal{H}_1$} (j);
        \path[->] (h) edge node[left] {no} node[right] {$\mathcal{H}_0$} (k);
        \draw[->, to path={-| (\tikztotarget)}] (k) edge node {} (a);
    \end{tikzpicture}
    }%
    \caption{Flowchart of the proposed MAST procedure. Input data are the daily infected. These noisy data are filtered to mitigate imperfections in data collection, randomness, and delays. Filtered daily positive $p_n$ are used to compute the growth rate $x_n$, which is used to compute the MAST statistic $T_n$. The statistic is then compared with the threshold $\chi$; if it is larger than the threshold, the outbreak is declared. Otherwise, the procedure continues to collect and process the data.}
    \label{fig:mast-block-diagram}
\end{figure}
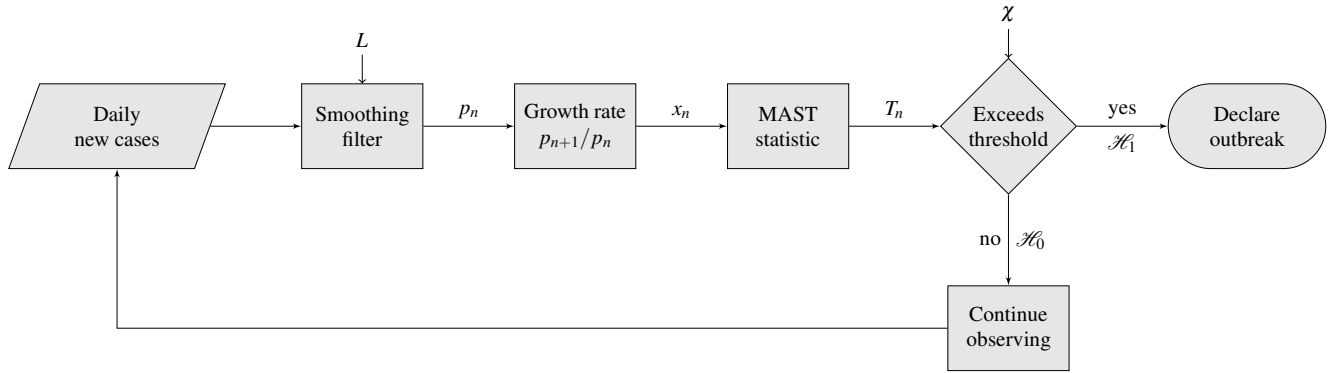

It is clear that managing an unprecedented pandemic event is a huge and a multifaceted problem, which can only be addressed by taking into account many different perspectives. It is also clear that the decision when to take pandemic countermeasures depends on a large number of societal factors. The contribution of this article is limited to the analysis of the pandemic strictly from a quickest-detection viewpoint and, from this perspective, we obtain useful insights and quantitative analyses. One evidence, as just pointed out, is that critical regimes of many nations began dramatically earlier than when countermeasures were taken. \textcolor{blue}{In this sense, we believe that the proposed decision support tool would be a key component of a command and control (C2) system that anticipates, and possibly reacts as soon as possible, to threats. Such a C2 system would be useful for both national health security and armed forces in the context of Chemical, Biological, Radiological and Nuclear (CBRN) defense.\cite{kumar2010chemical,ramesh2010triage}}

Aside from the above retrospective analysis, a major contribution of the MAST quickest detection tool developed in this paper consists of providing proactive decision support for detecting future waves of the COVID-19 outbreak, and the onset of future pandemics. In these cases, precise nation-dependent performance predictions, such as those given in Fig.~\ref{fig:operational}, cannot be available in advance because the curves have been obtained by exploiting estimates of the mean value of the growth rate (see e.g., the curve in magenta in Fig~\ref{fig:new-positives-and-growth-rate}), derived by forensic inspection of the data. However, good approximate performance bounds can be obtained by assuming constant growth rates $(1\pm\alpha)$ in the critical and controlled regime, respectively, for ``reasonable'' values of $\alpha$, for instance, the faster and the slower rates reported in Fig.~\ref{fig:new-positives-and-growth-rate} and used to compute the outermost black curves in Fig.~\ref{fig:operational}. The corresponding value of $\omega$ would be equal to the Kullback–Leibler information measure, which governs the performance of a clairvoyant Page's test that knows exactly the constant growth rate. 

Fig.~\ref{fig:mast-block-diagram} shows the flowchart of the MAST procedure. The filtering operation in Fig.~\ref{fig:mast-block-diagram} is important to mitigate gross errors and lack or delayed reporting of the input data \textcolor{blue}{(for instance, thousands of positive individuals from previous days are all reported on the current day, number of recovered individuals unavailable in the US data for a long time, etc.)} To compute $p_n$, the filtering operation used in this paper requires one to observe data samples beyond the current day $n$, which causes delays for on-line implementations. In these cases, alternative causal filtering strategies,\cite{oppenheim-SaS} \textcolor{blue}{such as the Savitzky-Golay filter~\cite{Schafer_2011}, would be} more appropriate. \textcolor{blue}{Another possibility to handle outliers could be the application of Huber's robust statistic.~\cite{huber2004robust} We leave such possible enhancements to future work.}

\begin{figure}

\hspace{8mm}
\begin{minipage}{0.30\textwidth}
    \centering
    \psfragfig[width=1\textwidth]{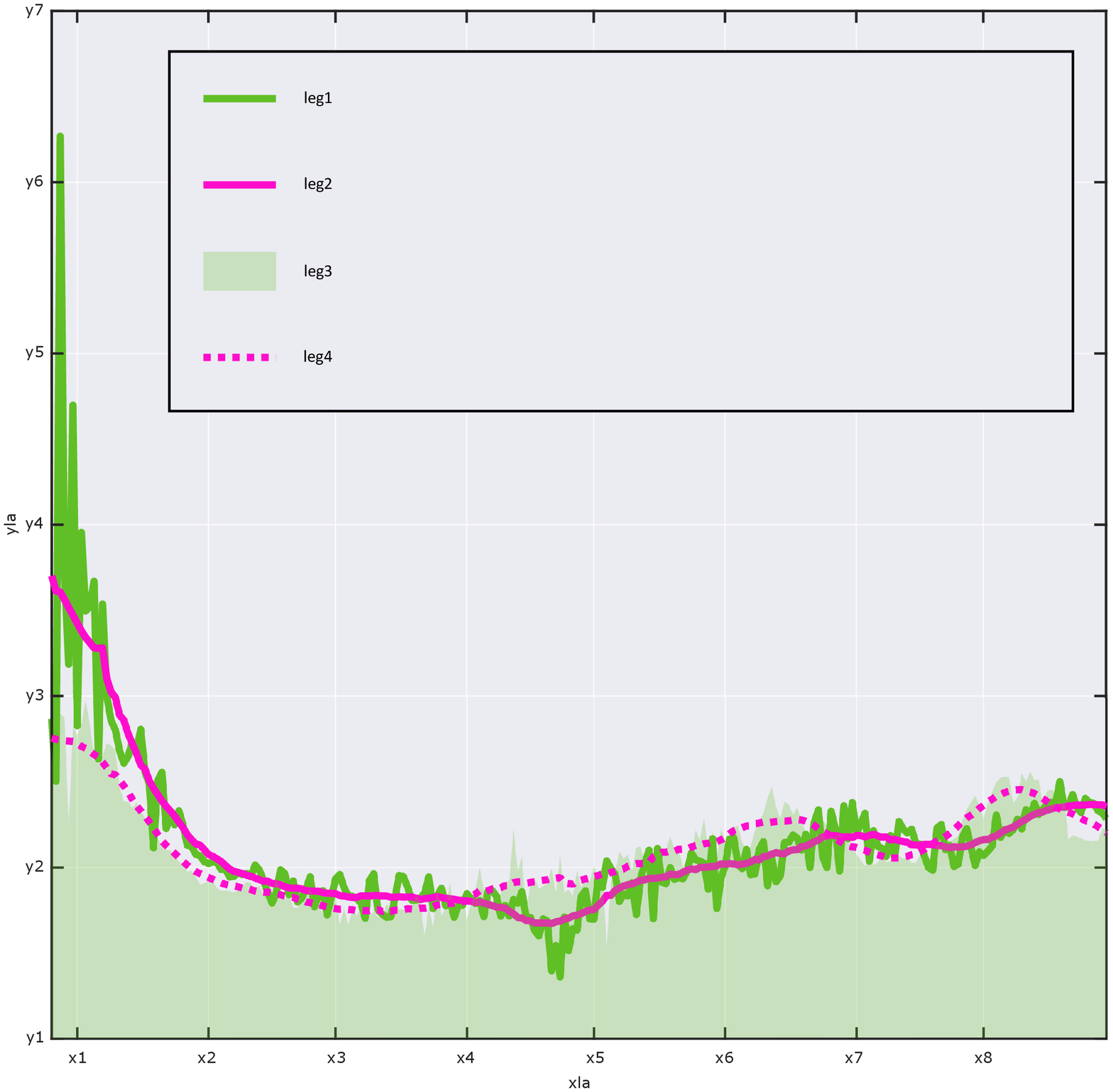} \\[4.5mm] \hspace{1.5mm} (a)
\end{minipage}%
\hspace{-3mm}
\begin{minipage}{0.70\textwidth}
    \centering
    \psfragfig[width=0.78\textwidth]{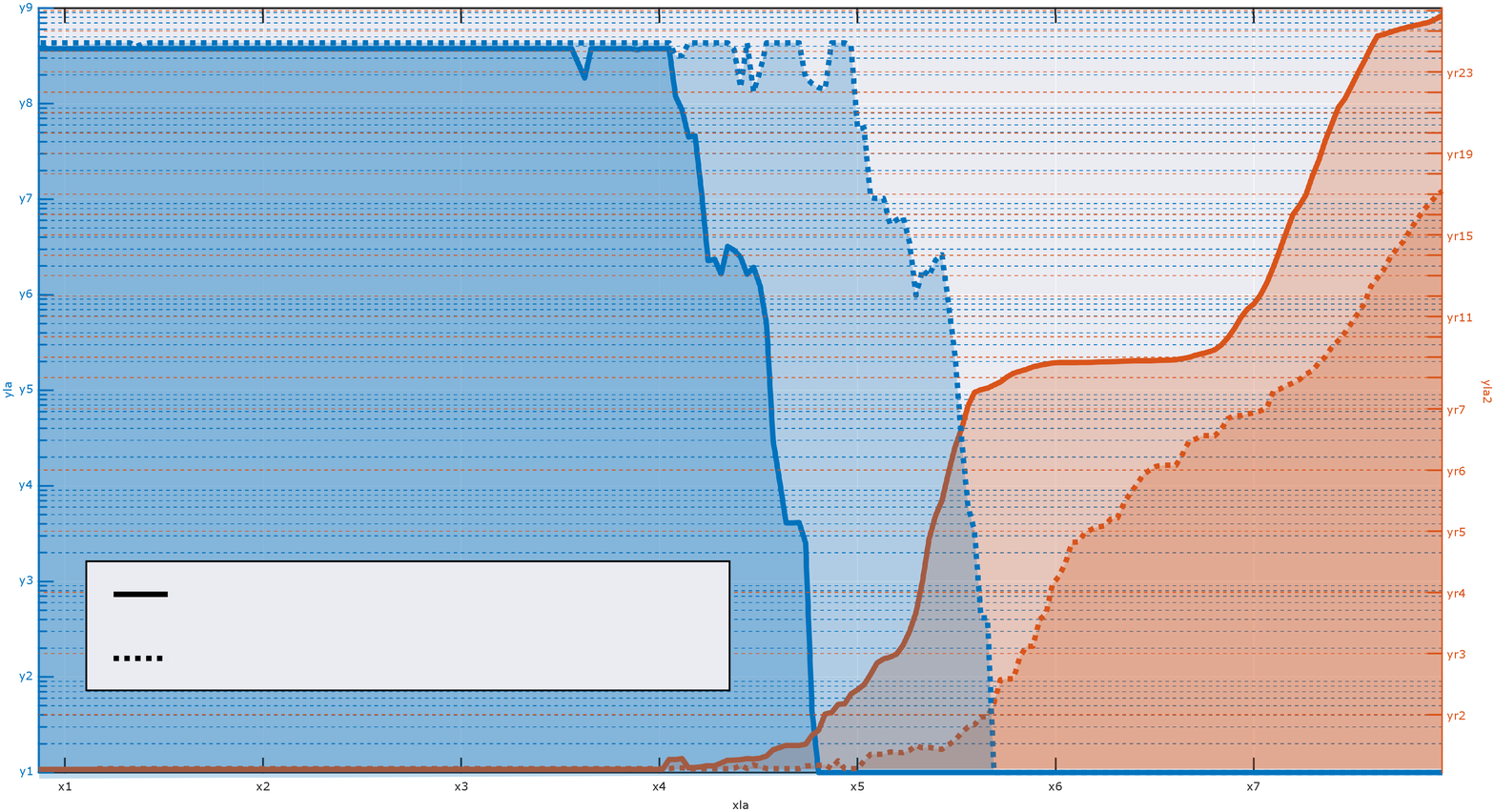} \\[4.5mm] \hspace{1.5mm} (b)
\end{minipage}%

\caption{(a) Growth rate of the
hospitalized individuals (green solid line) --- and its time-varying mean obtained through a moving average that uses a window of 21 days (magenta solid line) --- compared to the growth rate of the daily new positives individuals (green area) --- and its time-varying mean (magenta dotted line) --- in Italy since February 21, 2020. (b) Application of the MAST procedure on the growth rate sequence of the daily new positive individuals (solid lines, already shown in Fig.~\ref{fig:test-italy}) and on the growth rate sequence of the
hospitalized individuals (dotted lines).
On the left-side vertical axis we select a desired risk, e.g., $R = 10^{-4}$. Then, the blue curves indicate the stopping day (about July 18 if the growth rate sequence of the daily new positive individuals is used, and August 10 if the growth rate sequence of the
hospitalized individuals is used) corresponding to the selected value of risk. Finally, the red curves referred to the right-side vertical axis show the mean delay $\Delta$ corresponding to the selected risk $R$ (about 3 days if the growth rate sequence of the daily new positive individuals is used, and below 5 days if the growth rate sequence of the
hospitalized individuals is used). For clarity, note that the right-side scale for the delay is split into two linear ranges, for a better rendering of the small-$\Delta$ range.}

\label{fig:hospitalized-example-italy}

\end{figure}
\begin{figure}

\centering

\begin{minipage}{0.33\textwidth}
    \centering
    \psfragfig[width=.75\textwidth]{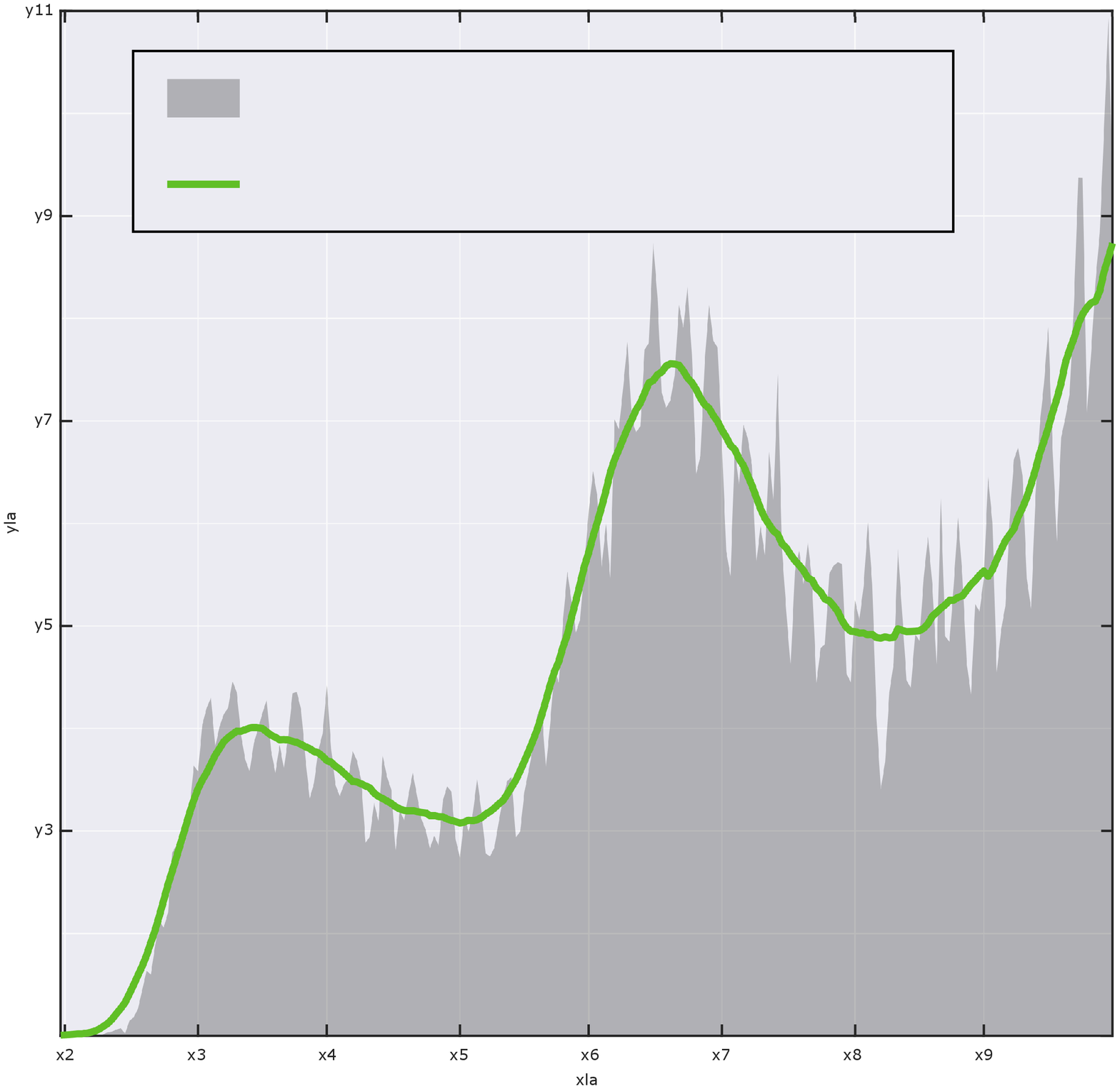} \\[4.5mm] \hspace{1.5mm} (a)
\end{minipage}%
\begin{minipage}{0.33\textwidth}
    \centering
    \psfragfig[width=0.75\textwidth]{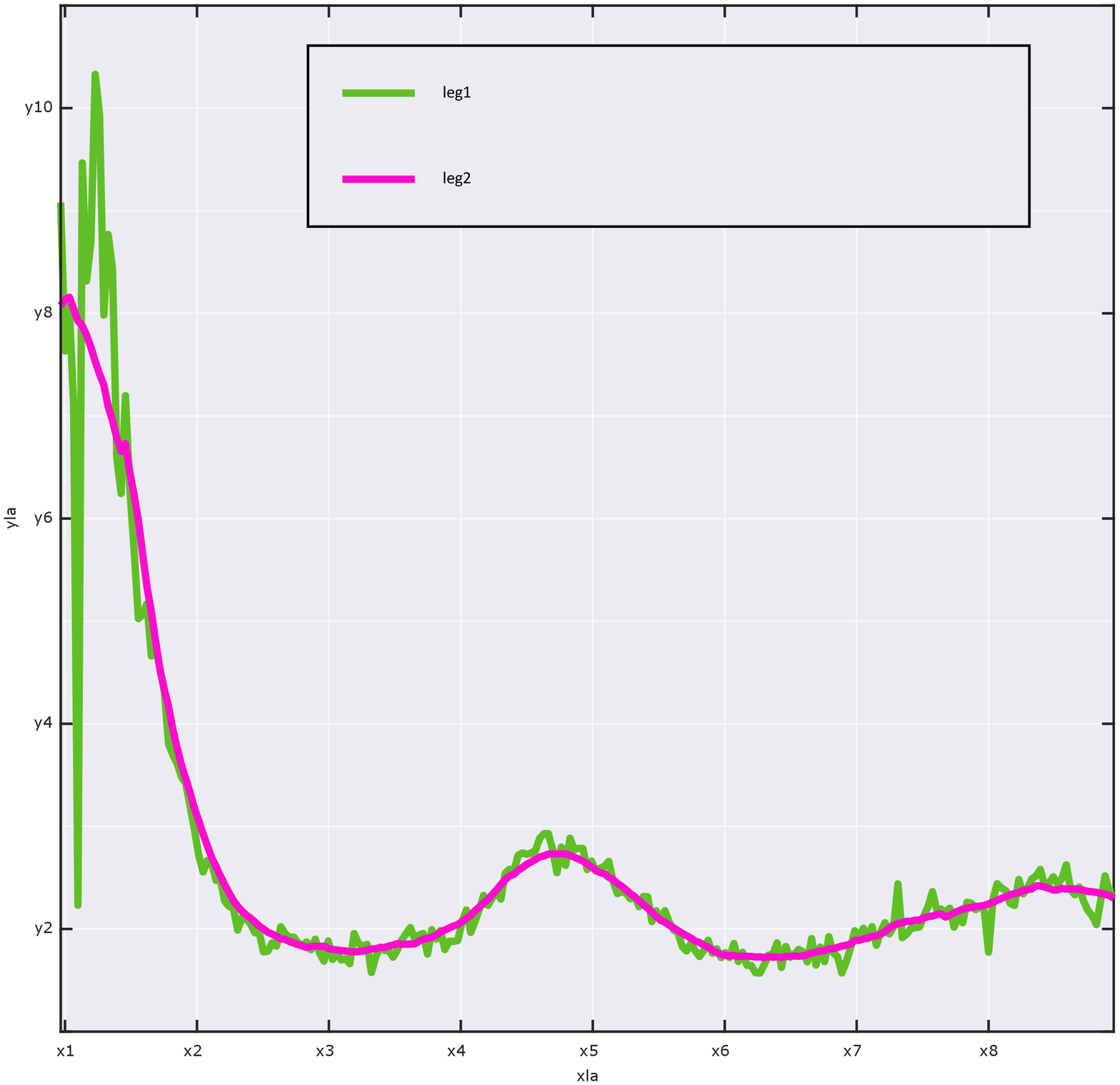} \\[4.5mm] \hspace{1.5mm} (b)
\end{minipage}%
\begin{minipage}{0.33\textwidth}
    \centering
    \psfragfig[width=0.79\textwidth]{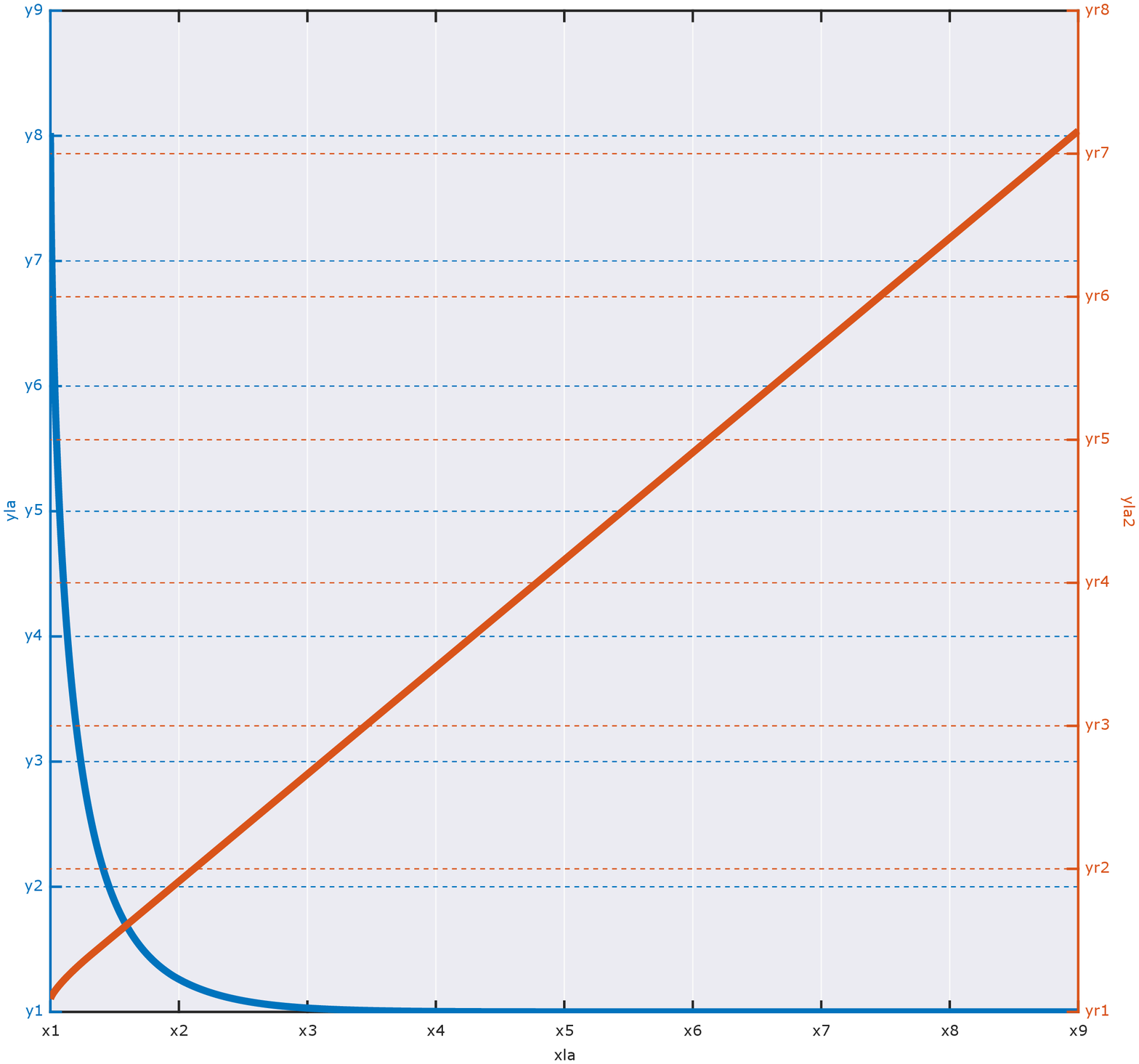} \\[4.5mm] \hspace{0mm} (c)
\end{minipage}%

\vspace{2mm}

\psfragfig[width=.75\textwidth]{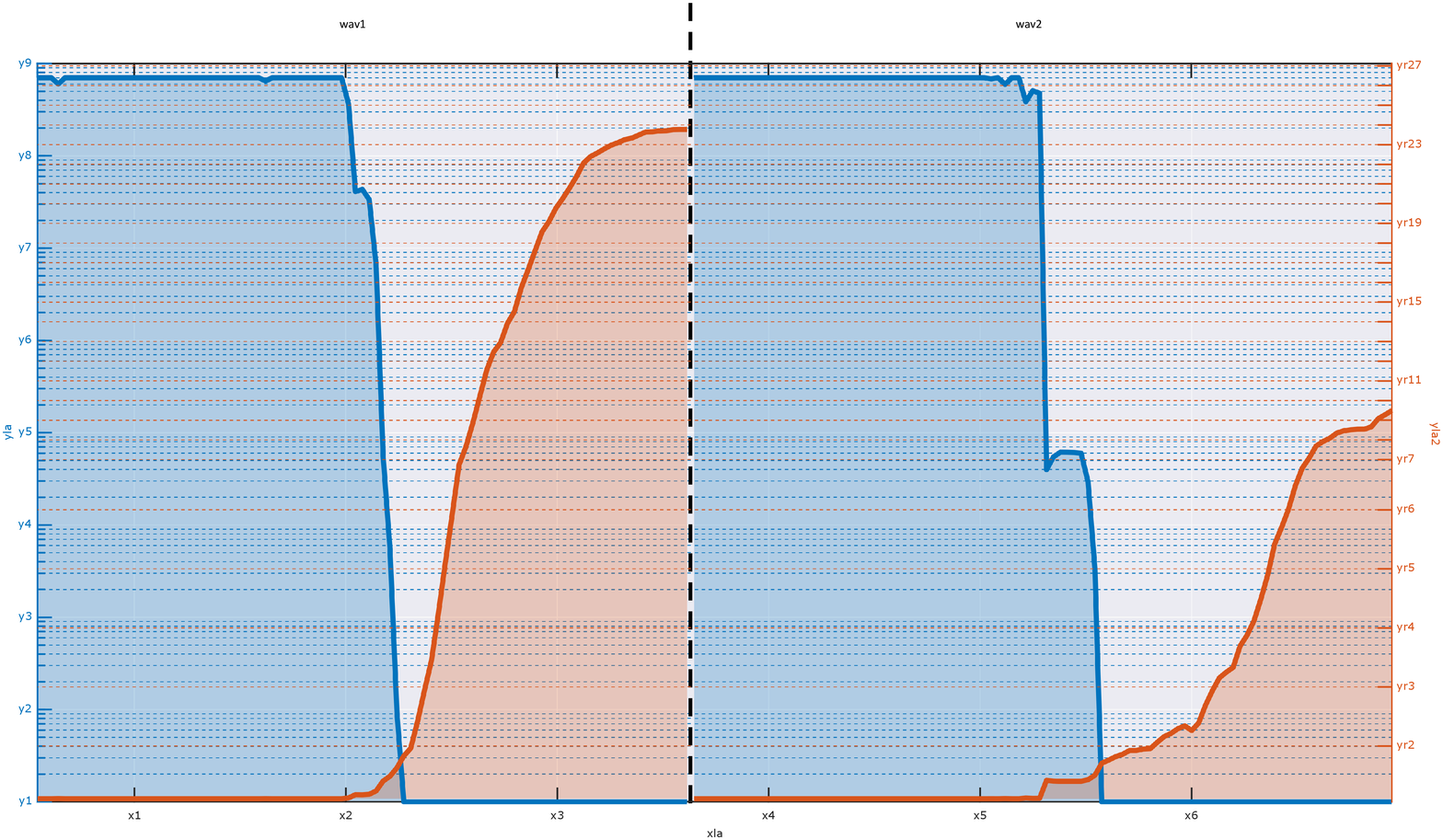} \\[4.5mm] \hspace{0mm} (d)

\caption{United States of America. (a) Daily new positive cases in the US since February 29, 2020, and its moving average obtained with a window of 21 days (green line).
(b) Growth rate of the epidemic computed from the averaged daily new positive cases (green line), and its time-varying mean obtained through a moving average that uses a window of 21 days (magenta line). (c) MAST performance, in terms of risk (left axis) versus threshold and mean delay (right axis) versus threshold, obtained from the US data of the COVID-19 pandemic. (d) Application of the MAST procedure on the COVID-19 pandemic data from US. Both the second wave (left side) and the third wave (right side) are analysed. For each wave, we select a desired risk on the left-side vertical axis, e.g., $R=10^{-4}$. Then, the blue curve indicates the stopping day (about June 6 for the first wave and September 10 for the third wave) corresponding to the selected value of risk. Finally, the red curve referred to the right-side vertical axis shows the mean delay~$\Delta$ corresponding to the selected risk~$R$ (approximately 4 days for both the second and the third waves). For clarity, note that the right-side scale for the delay is split into two linear ranges, for a better rendering of the small-$\Delta$ range.}

\label{fig:test-all-cas-US}

\end{figure}

The quickest-detection tool developed can also be applied to different time-series, other than the sequence of growth rate of daily new positives $\{x_n\}$, for instance to the hospitalized individuals addressed, for the Italian case, in Fig.~\ref{fig:hospitalized-example-italy}. 
The sequence of hospitalized individuals does not require the smoothing operation shown in the flowchart of Fig.~\ref{fig:mast-block-diagram}, because the collected data are less affected by gross errors. These data are also normally distributed with good accuracy, as confirmed by goodness-of-fit analysis (not reported). In Fig.~\ref{fig:hospitalized-example-italy} we see that the decision taken by using the hospitalized sequence is delayed as compared to that obtained by the sequence of daily new positives. This behaviour is intuitive because the hospitalized individuals are a subset of the positive ones, and a possible hospitalization follows the onset of clinical symptoms. 

\begin{figure}

\centering

\begin{minipage}{0.33\textwidth}
    \centering
    \psfragfig[width=.75\textwidth]{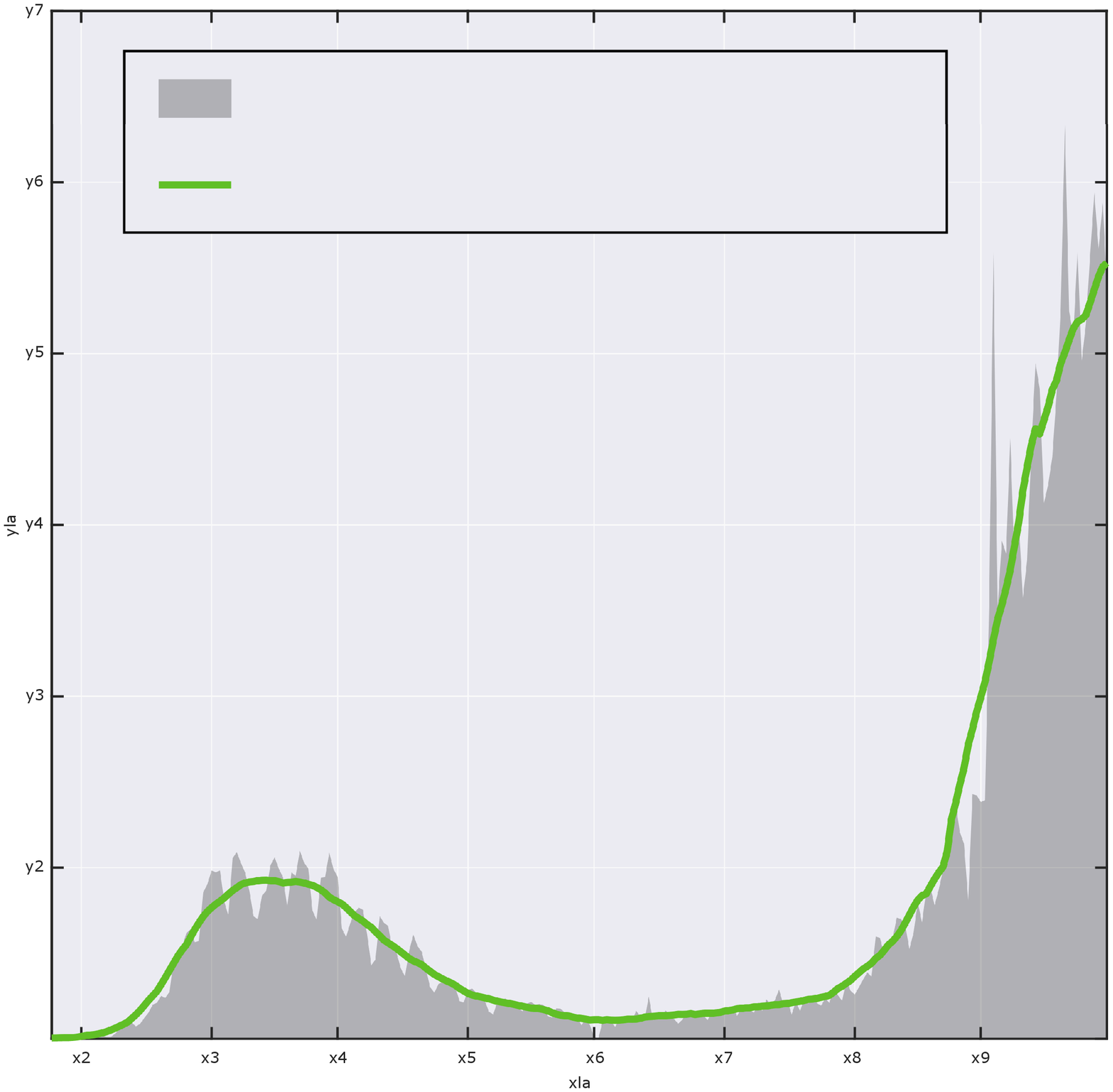} \\[4.5mm] \hspace{1.5mm} (a)
\end{minipage}%
\begin{minipage}{0.33\textwidth}
    \centering
    \psfragfig[width=0.75\textwidth]{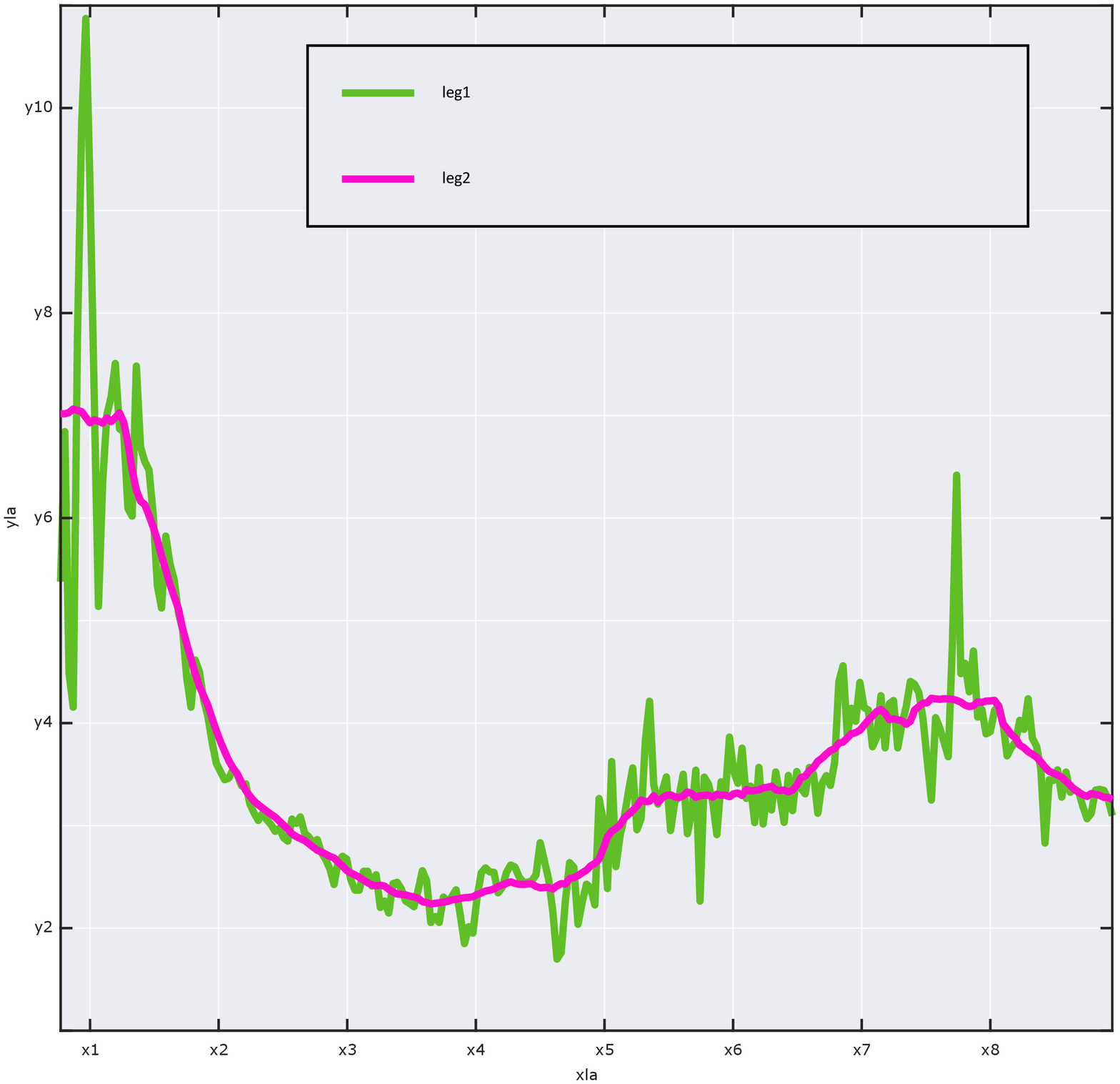} \\[4.5mm] \hspace{1.5mm} (b)
\end{minipage}%
\begin{minipage}{0.33\textwidth}
    \centering
    \psfragfig[width=0.79\textwidth]{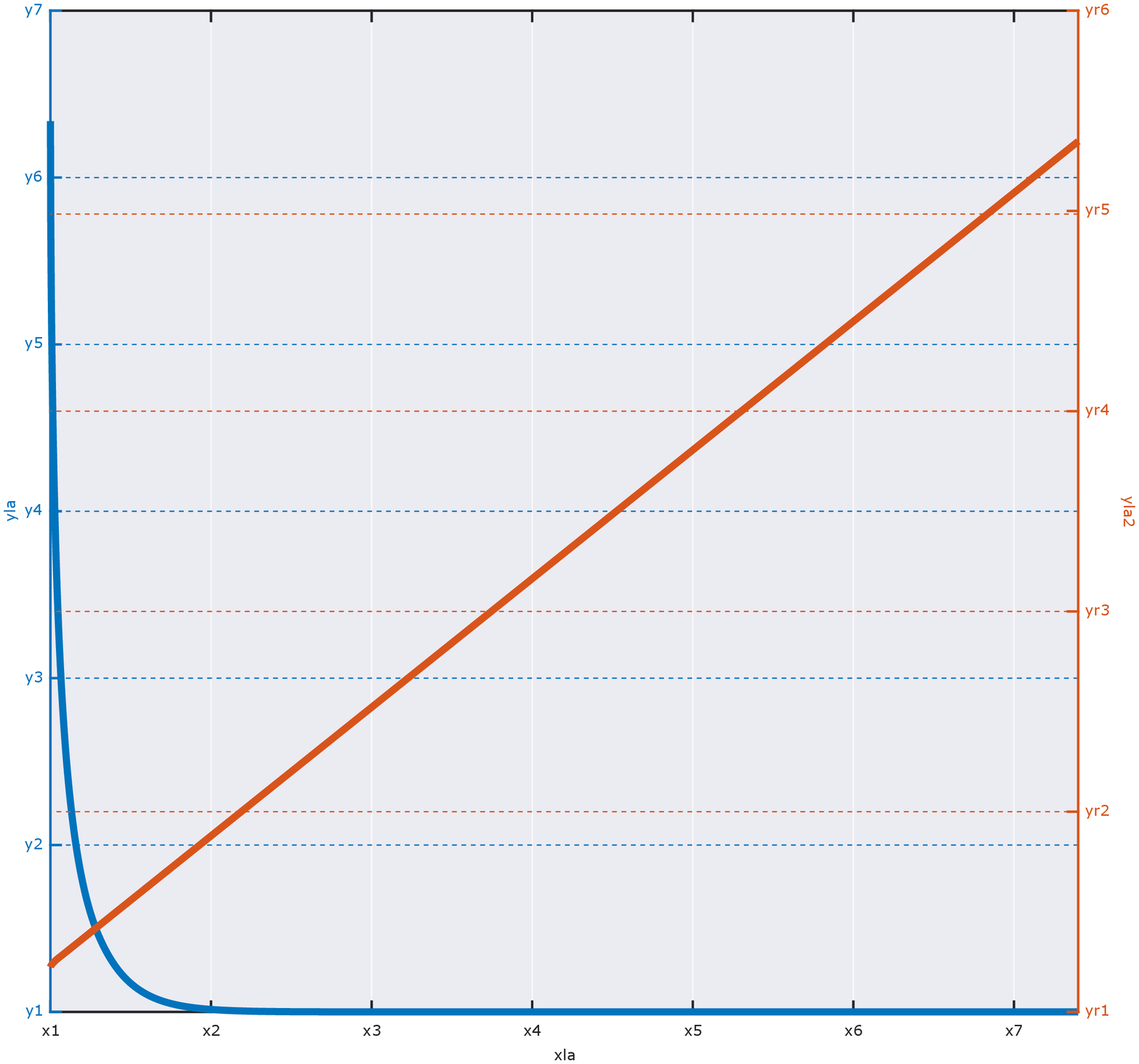} \\[4.5mm] \hspace{0mm} (c)
\end{minipage}%

\vspace{2mm}

\psfragfig[width=.75\textwidth]{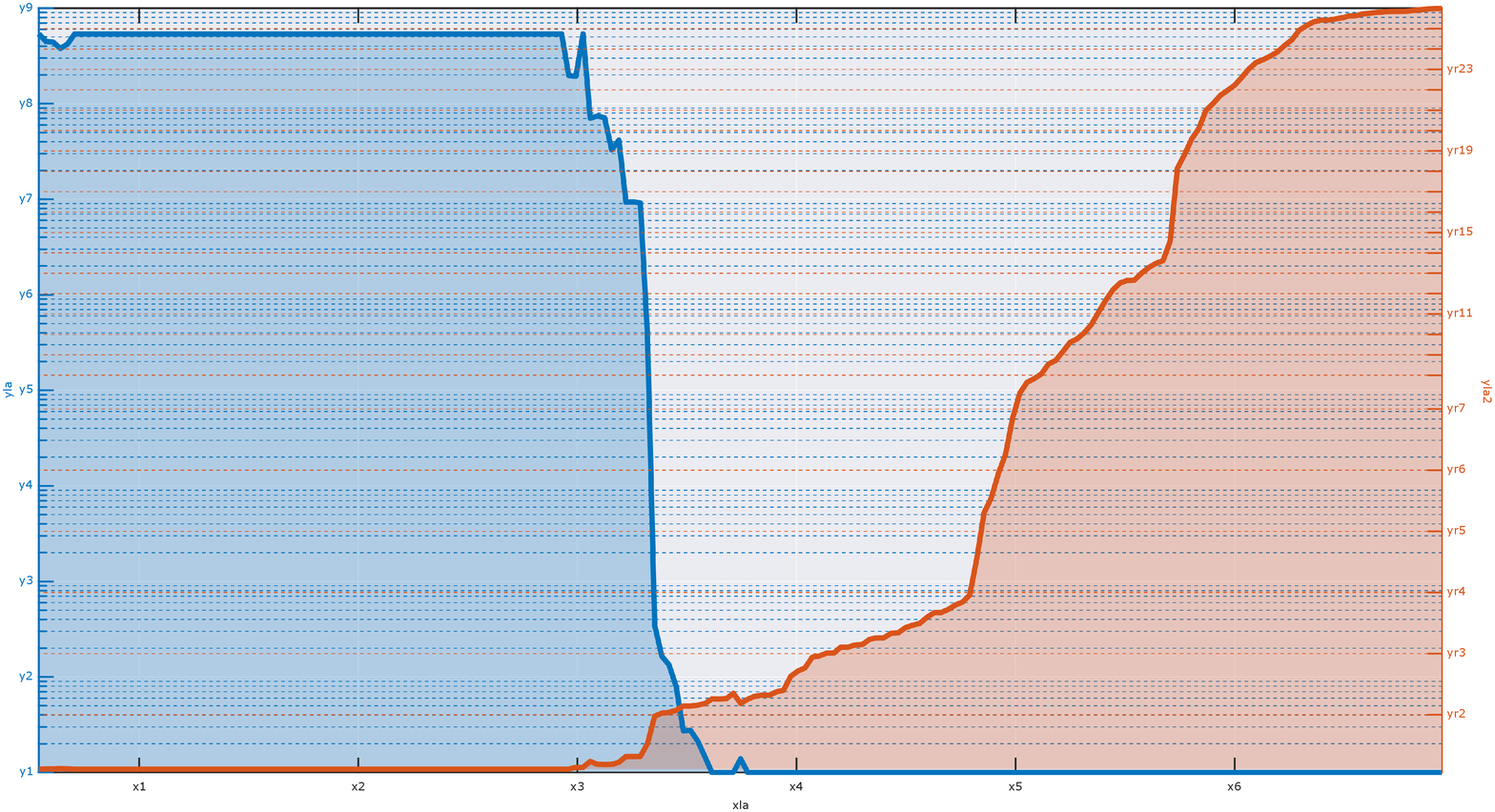} \\[4.5mm] \hspace{0mm} (d)

\caption{United Kingdom. (a) Daily new positive individuals in the UK since February 23, 2020, and its moving average obtained with a window of 21 days (green line).
(b) Growth rate of the epidemic computed from the averaged daily new positive cases (green line), and its time-varying mean obtained through a moving average that uses a window of 21 days (magenta line). (c) MAST performance, in terms of risk (left axis) versus threshold and mean delay (right axis) versus threshold, obtained from the UK data of the COVID-19 pandemic. (d) Application of the MAST procedure on the COVID-19 pandemic data from UK. On the left-side vertical axis, we select a desired risk, e.g., $R=10^{-4}$. Then, the blue curve indicates the stopping day (about July 11, in the example) corresponding to the selected value of risk. Finally, the red curve referred to the right-side vertical axis shows the mean delay~$\Delta$ corresponding to the selected risk~$R$ (below 6 days). For clarity, note that the right-side scale for the delay is split into two linear ranges, for a better rendering of the small-$\Delta$ range.}

\label{fig:test-all-cas-UK}

\end{figure}

\begin{figure}

\centering

\begin{minipage}{0.33\textwidth}
    \centering
    \psfragfig[width=.75\textwidth]{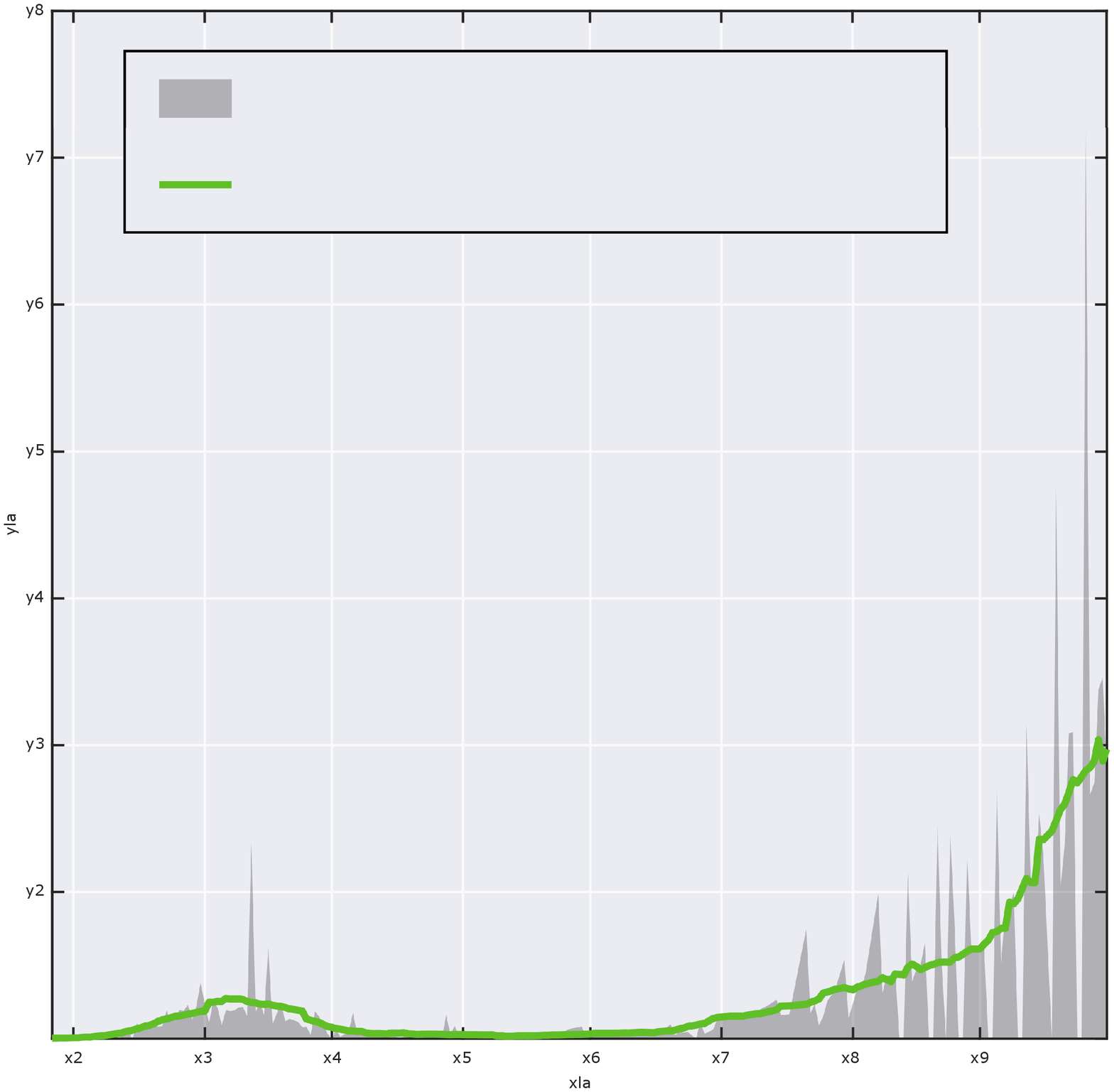} \\[4.5mm] \hspace{1.5mm} (a)
\end{minipage}%
\begin{minipage}{0.33\textwidth}
    \centering
    \psfragfig[width=0.75\textwidth]{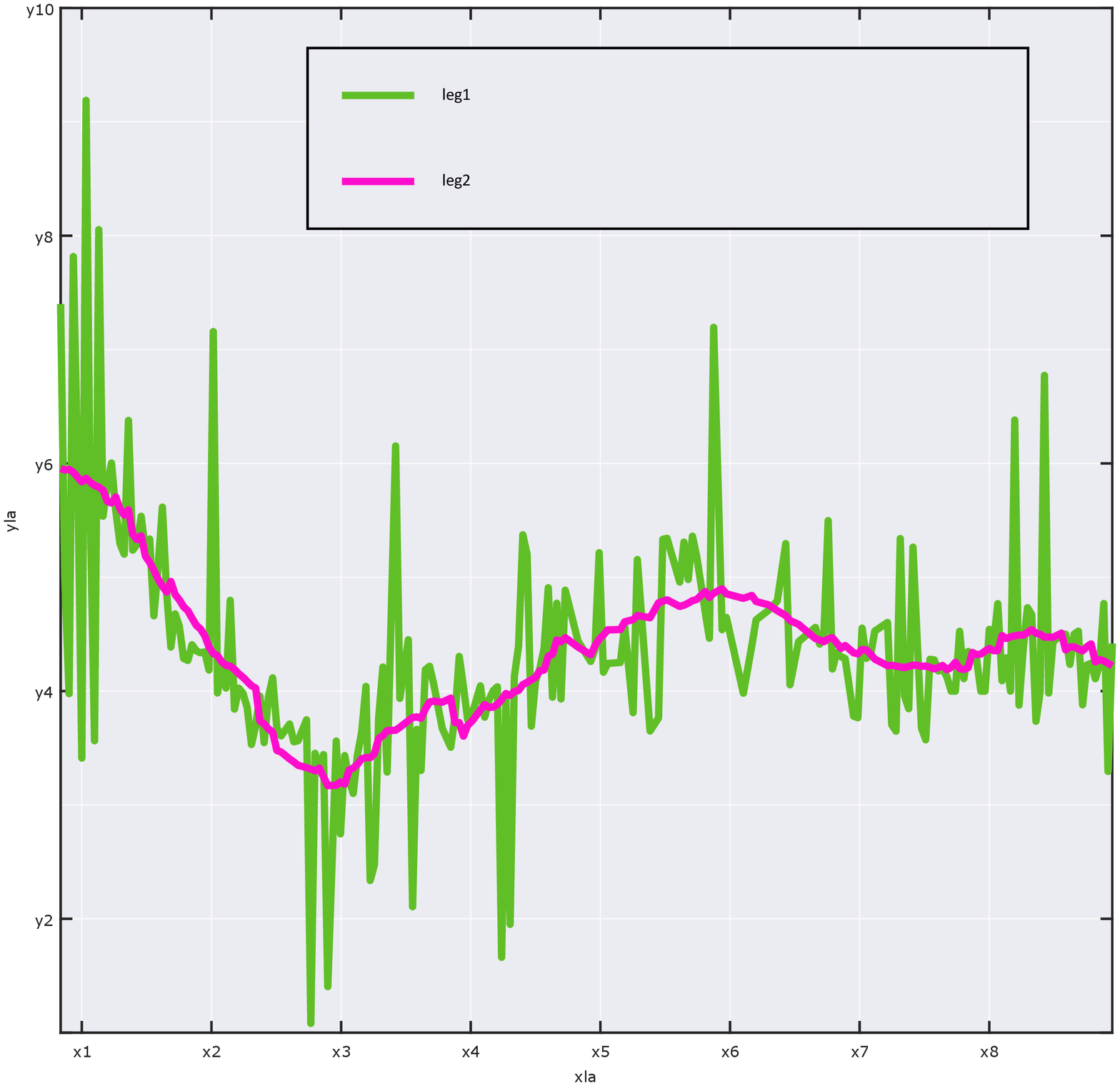} \\[4.5mm] \hspace{1.5mm} (b)
\end{minipage}%
\begin{minipage}{0.33\textwidth}
    \centering
    \psfragfig[width=0.79\textwidth]{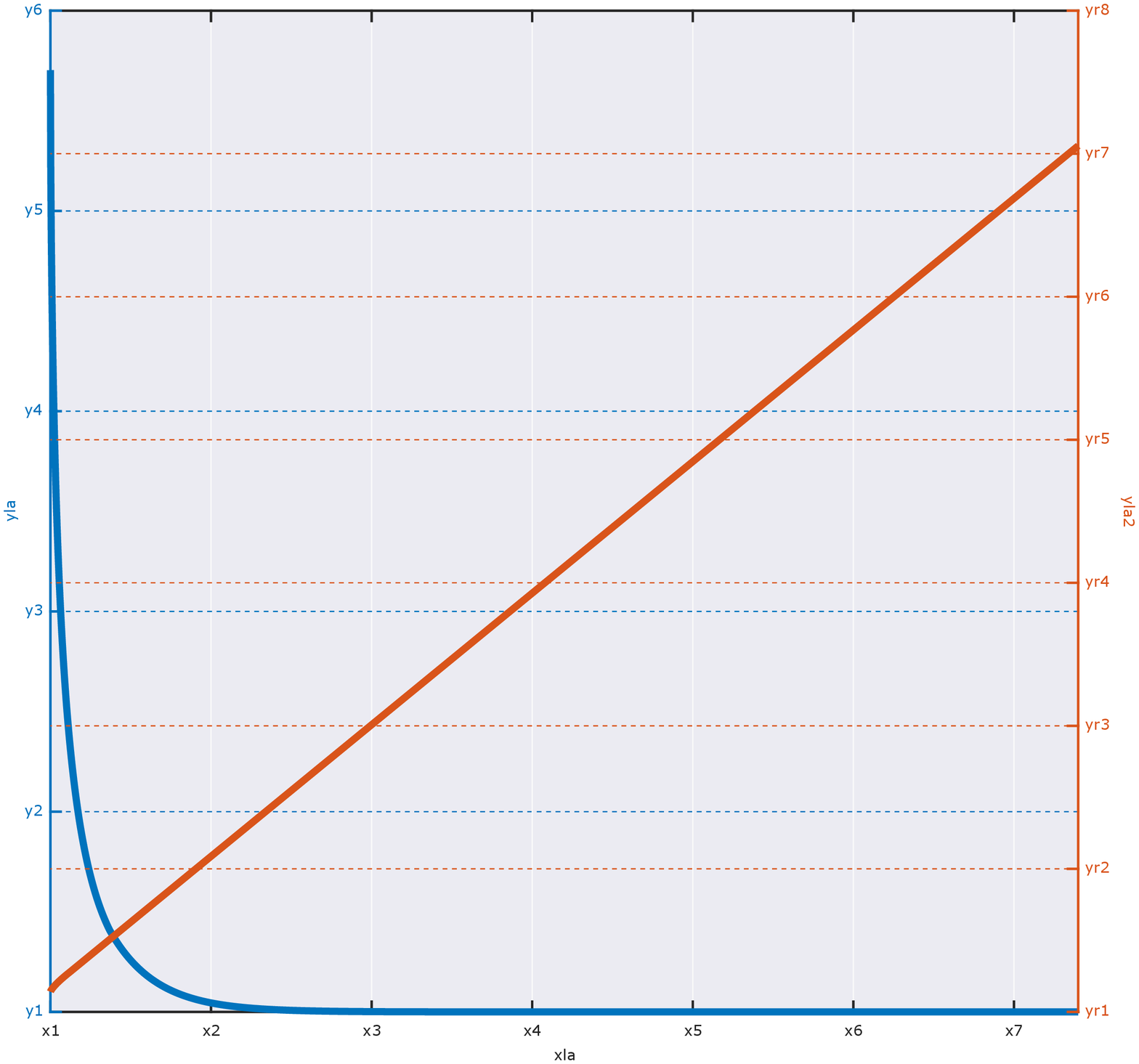} \\[4.5mm] \hspace{0mm} (c)
\end{minipage}%

\vspace{2mm}

\psfragfig[width=.75\textwidth]{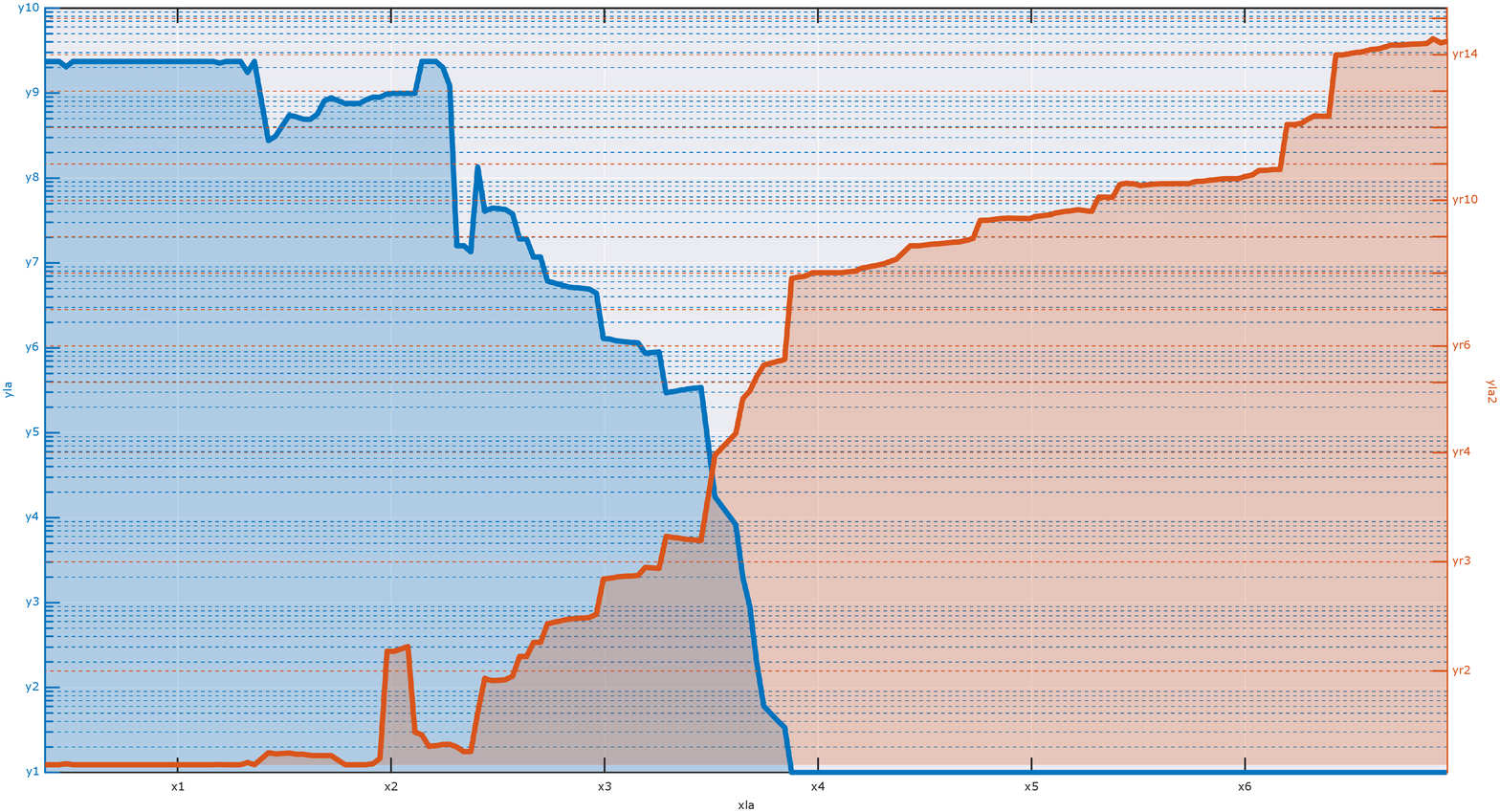} \\[4.5mm] \hspace{0mm} (d)

\caption{France. (a) Daily new positive individuals in France since February 25, 2020, and its moving average obtained with a window of 21 days (green line).
(b) Growth rate of the epidemic computed from the averaged daily new positive cases (green line), and its time-varying mean obtained through a moving average that uses a window of 21 days (magenta line). (c) MAST performance, in terms of risk (left axis) versus threshold and mean delay (right axis) versus threshold, obtained from the French data of the COVID-19 pandemic. (d) Application of the MAST procedure on the COVID-19 pandemic data from France. On the left-side vertical axis, we select a desired risk, e.g., $R=10^{-4}$. Then, the blue curve indicates the stopping day (about July 7, in the example) corresponding to the selected value of risk. Finally, the red curve referred to the right-side vertical axis shows the mean delay~$\Delta$ corresponding to the selected risk~$R$ (below 20 days). For clarity, note that the right-side scale for the delay is split into two linear ranges, for a better rendering of the small-$\Delta$ range.}

\label{fig:test-all-cas-France}

\end{figure}

\begin{figure}

\centering

\begin{minipage}{0.33\textwidth}
    \centering
    \psfragfig[width=.75\textwidth]{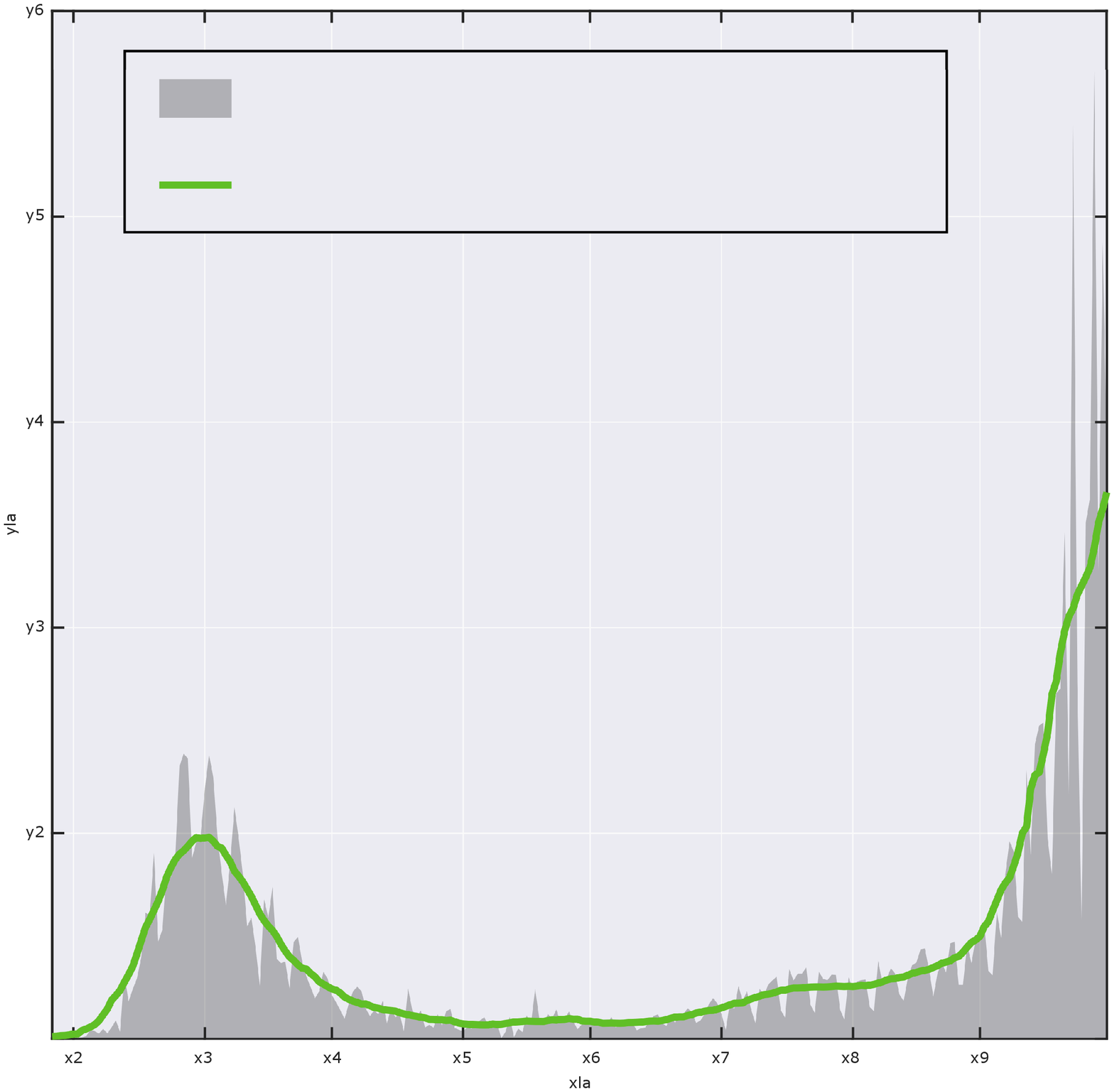} \\[4.5mm] \hspace{1.5mm} (a)
\end{minipage}%
\begin{minipage}{0.33\textwidth}
    \centering
    \psfragfig[width=0.75\textwidth]{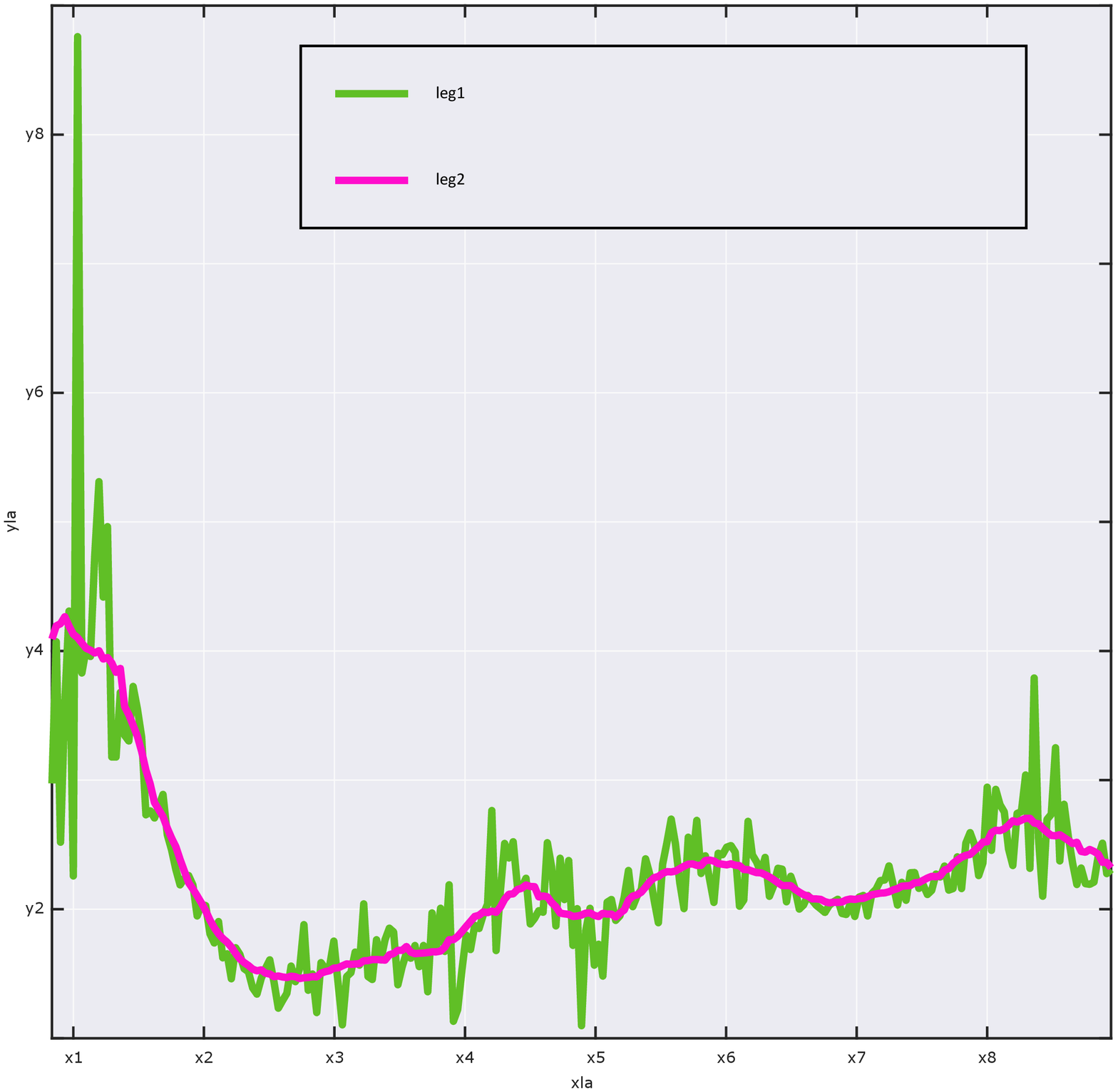} \\[4.5mm] \hspace{1.5mm} (b)
\end{minipage}%
\begin{minipage}{0.33\textwidth}
    \centering
    \psfragfig[width=0.79\textwidth]{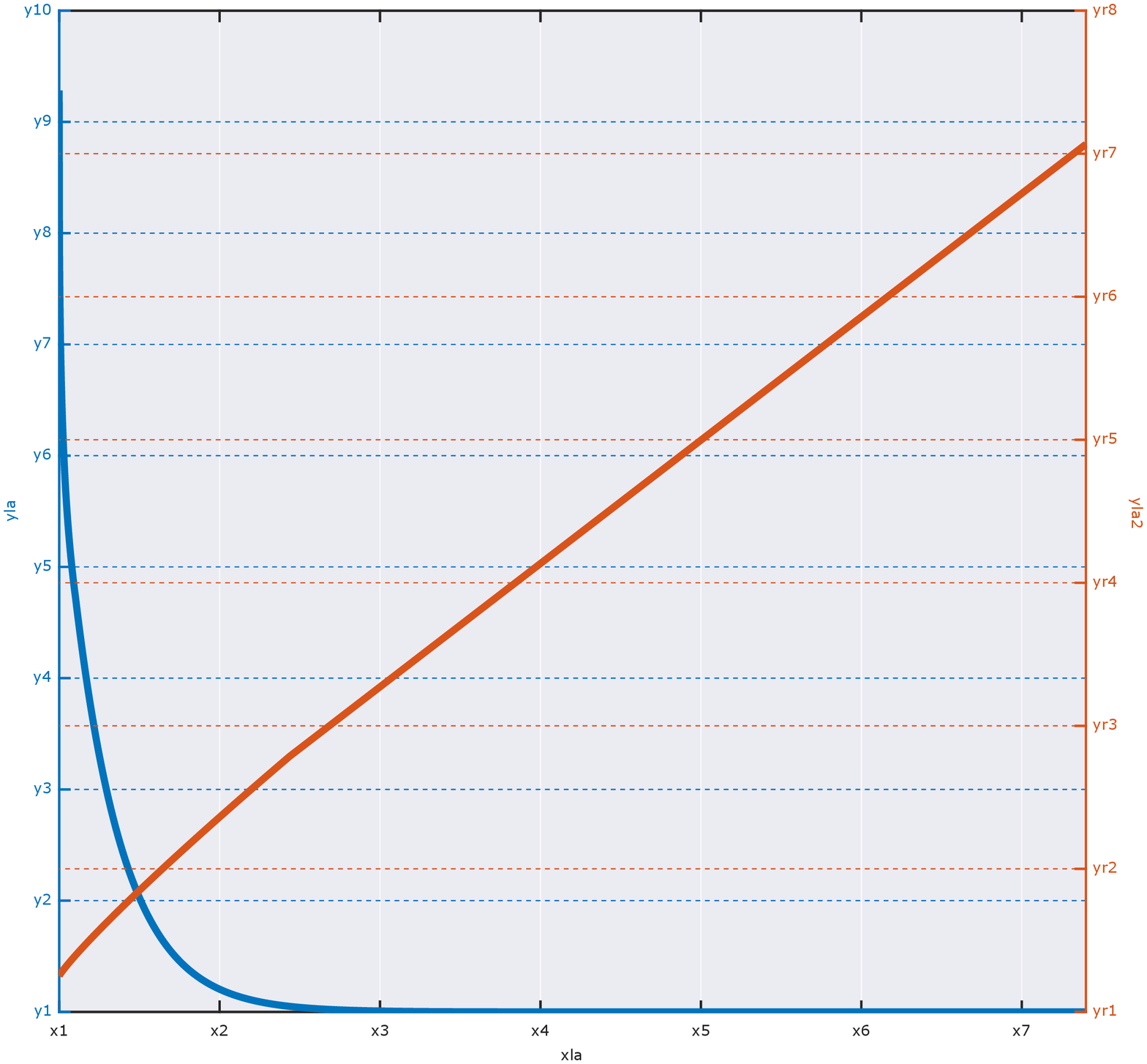} \\[4.5mm] \hspace{0mm} (c)
\end{minipage}%

\vspace{2mm}

\psfragfig[width=.75\textwidth]{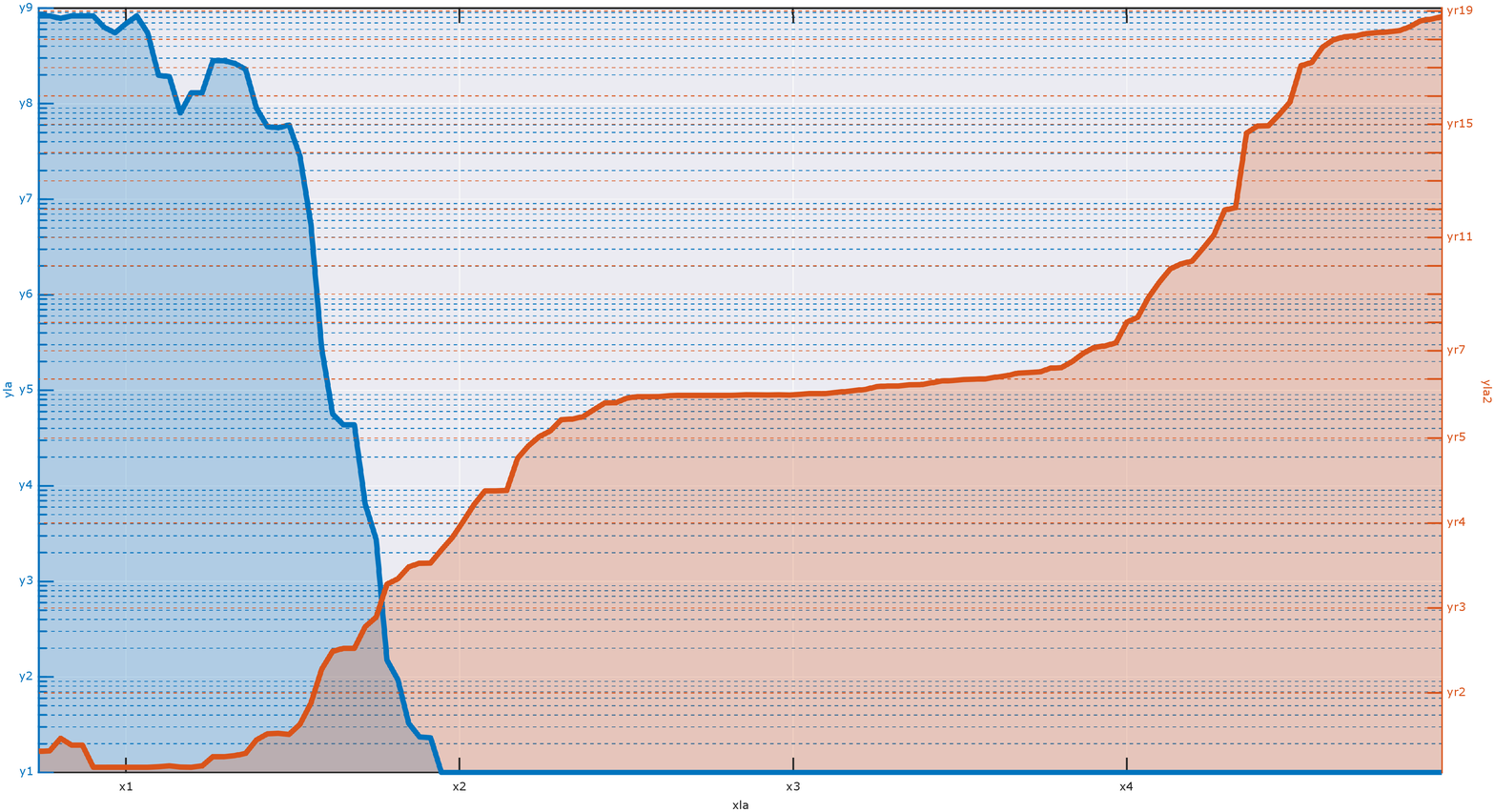} \\[4.5mm] \hspace{0mm} (d)

\caption{Germany. (a) Daily new positive individuals in Germany since February 25, 2020, and its moving average obtained with a window of 21 days (green line).
(b) Growth rate of the epidemic computed from the averaged daily new positive cases (green line), and its time-varying mean obtained through a moving average that uses a window of 21 days (magenta line). (c) MAST performance, in terms of risk (left axis) versus threshold and mean delay (right axis) versus threshold, obtained from the German data of the COVID-19 pandemic. (d) Application of the MAST procedure on the COVID-19 pandemic data from Germany. On the left-side vertical axis, we select a desired risk, e.g., $R=10^{-4}$. Then, the blue curve indicates the stopping day (about July 19, in the example) corresponding to the selected value of risk. Finally, the red curve referred to the right-side vertical axis shows the mean delay~$\Delta$ corresponding to the selected risk~$R$ (below 13 days). For clarity, note that the right-side scale for the delay is split into two linear ranges, for a better rendering of the small-$\Delta$ range.}

\label{fig:test-all-cas-Germany}

\end{figure}

\textcolor{blue}{Public opinion soundings suggest increased concern about new SARS-CoV-2 variants,\cite{alteri2021genomic} especially related to their higher fatality rate,~\cite{becerra2020sars} spreading velocity,\cite{kirby2021new,Poor2020} and the possibility that approved vaccines might be less effective against them.\cite{conti2021british,kemp2020neutralising,Kupferschmidt329} 
The proposed quickest detection tool is useful also in the presence of variants of the coronavirus, which would affect the growth rate; indeed, variants were already present in the United Kingdom during the second wave, reported here in  Fig.~\ref{fig:test-all-cas-UK}. It is worth noting here that the MAST statistic does not require knowledge of the time-varying mean value sequence of the growth rate.}

The possibility of processing different sequences of data opens the way to the design of more sophisticated decision rules based on joint processing of multiple time-series. In addition, the MAST procedure can easily accommodate different definitions of critical regime. For instance, if a pandemic growth at rate $(1+\alpha^*)$ is considered acceptable,
for some $0<\alpha^*\ll1$, the critical regime can be characterized by rates exceeding~$(1+\alpha^*)$, rather than~1. 
The necessary modifications to the MAST statistic are straightforward.
\textcolor{blue}{We also envision that considering both the passage from a controlled to a critical regime \emph{and vice versa} can be addressed by minor modifications to the MAST procedure.\cite{soldi2021commag,MarSay-submitted}}

\textcolor{blue}{Special attention is given in the literature to the evaluation of underreporting and undertesting of COVID-19 cases.~\cite{Li489,LAU2021110} The mortality rate is used as the main indicator to evaluate the extent of underreporting and underdetection of COVID-19 cases.~\cite{LAU2021110} However, in our context, the evaluation of underreporting cases would not be beneficial in terms of quickest detection, as it is provided by an estimation procedure that uses the same data processed by the quickest detection. As already mentioned, all the available  data (hospitalized individuals, daily deaths, daily number of tests etc.) could be used jointly and thus improve the detection capability and reliability of the approach.}

An extended analysis of COVID-19 infection data from more countries than those covered in this paper is available on the web.~\cite{WEBSITE}
%
We hope that in the near future the publicly available data can be: (i) more reliable so as to mitigate bias effects due to, e.g., false positives, contrasting multiple test outcomes for the same individual, markedly different contagion incidence in close geographical areas, etc.; and (ii) released with finer granularity so as to allow for analyses stratified by population age, comorbidity, etc. These aspects are also relevant \textcolor{blue}{for effective vaccination policies.}

\section*{Methods}

The observation model used in this paper can be formally obtained by replacing the constant growth rate $(1+\alpha)$ appearing in Eq.~\eqref{eq:withalpha} by the sequence of random variables $x_1,x_2,\dots$, yielding 
\begin{align}
    p_{n+1}=p_1 \prod_{k=1}^n x_k, \quad n=1,2,\dots
    \label{eq:withx}
\end{align}
where $p_n$ is the number of new cases on day $n$, while $x_n=p_{n+1}/p_{n}$, a time-series that makes explicit the growth rate that we seek to coopt. The model in Eq.~\eqref{eq:withx} is validated empirically. In particular, in the accompanying Supplementary Information, we elaborate on the classic susceptible-infected-recovered (SIR) compartmental epidemic 
model to motivate the usage of the sequence $\{x_n\}$ as observable process for quickest detection. 
Using the model in Eq.~\eqref{eq:withx}, we assume to have available the sequence of daily new positives, for a certain region of interest; the analysis presented here relies on the data provided by the Johns Hopkins University.~\cite{covid-19-JHU} Referring for instance to Fig.~\ref{fig:new-positives-and-growth-rate}, such a sequence is shown in gray in the left panel. To address gross errors, missing values and delays in reporting the data, the 
sequence is smoothed by a moving average filter with uniform weights (the moving average filter of length $L$ is always assumed to have equal $1/L$ weights). The smoothed sequence $\{p_n\}$ so obtained is shown in green in Fig.~\ref{fig:new-positives-and-growth-rate}(a). The growth rate process used as observable is then computed as $\{x_n=p_{n+1}/p_{n}\}$, and is represented by the green curve in Fig.~\ref{fig:new-positives-and-growth-rate}(b). \textcolor{blue}{The length of the filter is selected to $L=21$ as to obtain a convenient mitigation of the effects of gross errors in all analyzed countries. After such a pre-processing, we implement 
Kolmogorov-Smirnov tests to check data Gaussianity, see details in the Supplementary Information.}

The time-varying statistical mean $\{\mu_n\}$ of the sequence can be estimated by low-pass filtering of the sequence $\{x_n\}$, and for this we use again a moving average filter of length $L=21$ days.
By subtracting from each $x_n$ the estimated mean value $\widehat \mu_n$, that is, the curve in magenta in Fig.~\ref{fig:new-positives-and-growth-rate}(b), one obtains the sequence ${x_n-\widehat \mu_n}$.
Statistical analysis conducted by Kolmogorov-Smirnov goodness-of-fit test\cite{feller1948} reveals that ${x_n-\widehat \mu_n}$, for each $n=1,2,\dots$, can be modeled by a zero-mean Gaussian random variable with (country-dependent) standard deviation $\sigma \ll 1$. Since $\mu_n$ is close to unity, we see that $x_k < 0$ with negligible probability, hence the Gaussian approximation should not be problematic.

By observing the sequence $\{x_n\}$ as time $n$ elapses, we want to detect the passage from the controlled regime ${\cal H}_0$, to the critical regime ${\cal H}_1$, when there is the 
exponential growth. In the controlled regime, the mean value of the random variable~$x_n$ is below one, i.e., $\mu_n=\mu_{0,n} \le 1$, meaning that the number of new positives remains approximately stable or decreases. Conversely, in the critical regime, the mean value is greater than unity, $\mu_n =\mu_{1,n} > 1$, which implies an explosion of daily new positives in the long run. Formally:
\beqa
\textnormal{controlled regime } {\cal H}_0 &\Rightarrow &  \; \; x_n \sim \N(\mu_{0,n}, \sigma),  \qquad \mu_{0,n} \le 1, \label{testcontrolled}\\
\textnormal{critical regime } {\cal H}_1  &\Rightarrow & \; \; x_n \sim \N(\mu_{1,n}, \sigma),  \qquad  \mu_{1,n} > 1 \label{testcritical}. 
\eeqa
We assume that the $x_k$'s are mutually independent under either regime. \textcolor{blue}{As we show in the accompanying Supplementary Information, slight deviations from the condition of perfect independence do not significantly affect the results of the proposed MAST.}
Different nations are characterized by different values of $\sigma$, hence for each country $\sigma$ is assumed known, because in practice it can be estimated on-line from the data. Conversely, the quantities $\{\mu_{0,n}\}$ and $\{\mu_{1,n}\}$ are modeled as deterministic but \emph{unknown} sequences. 

To detect the change of regime, we rely on the Generalized Likelihood Ratio Test (GLRT) approach,~\cite{poorbook,Lehmann-testing,shao} a milestone of decision theory in scenarios where the statistical distributions of the data contain unknown parameters --- in our case, the sequences of mean values $\{\mu_{0,n}\}$ and $\{\mu_{1,n}\}$, and the time at which the passage of regime occurs. If we assume the sequences of mean values in the two regimes to be constant and known, say $\mu_{0,n}=1-\alpha$ and $\mu_{1,n}=1+\alpha$ (cf. equation~\eqref{eq:withalpha}), the GLRT solution to the quickest-detection problem would be the celebrated Page's test: compare to an appropriate threshold the CUSUM statistic\cite{page,poor-book-quickest,basseville-book}
\begin{align}
\hspace*{-10pt}\begin{cases}
Q_0=0, \\
Q_{n}=\max \bigg \{ 0, Q_{n-1} + \frac{2 \alpha \, (x_n-1) }{\sigma^2}  \bigg \}, \quad  n \ge 1.
\end{cases} 
\label{eq:page}
\end{align}
In the presence of unknown and time-varying sequences complying with the constraints on $\mu_{0,n}$ and $\mu_{1,n}$, shown in~\eqref{testcontrolled}-\eqref{testcritical},
the GLRT solution to the quickest-detection problem amounts to compare the MAST statistic~$T_n$ to a threshold~$\chi$:~\cite{Letter-arXiv} 
\begin{align}
\hspace*{-10pt}\begin{cases}
T_0=0, \\
T_{n}=\max \bigg \{ 0, T_{n-1} + \frac{(x_n-1)^2 \sign(x_n-1)}{2 \sigma^2} \bigg \}, \quad n \ge 1,
\end{cases} 
\label{eq:pagesquare}
\end{align}
and declaring the change of regime at the first occurrence of the threshold crossing. 
It is worth noting that the MAST statistic is formally obtained by replacing the unknown value of $\alpha$ appearing in the CUSUM statistic, with the estimate $\widehat \alpha_n=|x_n-1|$ (constant factors can be incorporated in the threshold). The reader educated in detection theory will recognize the analogy with the energy detector arising in testing the presence of an unknown time-varying deterministic signal buried in Gaussian noise.~\cite{kaydetection}

The threshold $\chi$ employed in the MAST procedure is selected to trade-off decision delay $\Delta$ and risk $R$, two quantities that are defined as follows.
The mean delay $\Delta$ is the mean value of the difference between the time at which the MAST statistic crosses the threshold~$\chi$ and the time of passage from the controlled to the critical regime. 
The risk $R$ is defined as reciprocal of the mean time between successive false alarms (in fact, ``false alarm probability'' is perhaps a more familiar jargon to readers with background in detection theory), where the false alarm is defined as a threshold crossing during the controlled regime (i.e., the regime in which there is no need for intervention). In this paper, the functions $R(\chi)$ and $\Delta(\chi)$ are obtained by standard Monte Carlo simulations~\cite{kaydetection, Allen2017, Lehmann-testing} for relatively small values of $\chi$, see e.g., Fig.~\ref{fig:new-positives-and-growth-rate}(c).
We found that the first function is essentially exponential, and the second is essentially linear, consistent with known expressions for the Page's test.~\cite{basseville-book,poor-book-quickest} This allows us to extrapolate the behaviour of the two functions for values of $\chi$ that would be problematic to obtain
from real data or by computer experiments, such as those used in Fig.~\ref{fig:test-italy} and similar figures for other nations.


	\clearpage
    \bibliography{mybib_upd_clean}
	
\section*{Acknowledgements}
PW was supported by AFOSR under contract FA9500-18-1-0463.

	
\section*{Author contributions statement}
SM developed the approach and supervised the work. PB, LMM, and DG did most of the analysis. DG prepared the figures. LMM prepared the website. PW and KP supervised the work. All authors contributed in writing the manuscript.

\section*{Additional information}
\urlstyle{same}
A visualization of MAST performance on updated COVID-19 infection data from an extended set of countries is available at \url{https://covid-mast.github.io}.

\section*{Competing interest}
The authors declare no competing interests.

	
	


\cleardoublepage
\appendix
\section*{Supplementary Information}

\subsection*{Observation Model}

\subsubsection*{Classical Deterministic SIR}
\label{sec:SIR}

Let us start, for the sake of simplicity, from the classical Susceptible-Infectious-Recovered (SIR) model of epidemic spread; a similar discussion holds for more complicated compartmental models. There are $P$ individuals (for concreteness, this may be the population of a state or a country), grouped into three classes, or \textit{compartments}: susceptible $S(t)$, infected $I(t)$, and removed (or recovered) $R(t)$. Let $s(t)=S(t)/P$, $i(t)=I(t)/P$ and $r(t)=R(t)/P$, denote the corresponding fractions of individuals. Each susceptible individual infects, on the average, $\beta$ randomly-chosen other individuals per unit of time. Each infected individual will then recover (or, unfortunately, pass away) at a constant average rate $\gamma$.
The popular SIR model~\cite{KerMckWal:J27,Allen2017,AdaptiveBayesian} is formalized mathematically as follows
\begin{align} \label{eq:SIR}
\begin{cases}
\frac{ds(t)}{dt} &= - \beta \, s(t) i(t), \\
\frac{di(t)}{dt} &= \beta \, s(t) i(t) - \gamma \, i(t), \\
\frac{dr(t)}{dt} &=  \gamma \, i(t),
\end{cases} 
\end{align}
with initial conditions $r(0)=0$, $s(0)=1-i(0)$; $i(0)$ is a (fairly small) fraction of the total population that gives rise to the spread of the infection.
Note that one of the three equations in~\eqref{eq:SIR} is redundant, in view of the conservation constraint $r(t)+s(t)+i(t)=1$.
Dividing the first of~\eqref{eq:SIR} by $s(t)$ and solving for $i(t)$ from the third equation, one gets 
$d \log s(t)= - \frac \beta \gamma \, d r(t)$. The quantity $\beta/\gamma$ goes also under the name of \emph{contact number} and is a combined characteristic of the population and the disease. Also, using $i(t)=1-r(t)-s(t)$ in the third equation, we eliminate $i(t)$ and arrive at
\begin{align}
\begin{cases} \label{eq:SIRsol}
s(t)&= s(0) e^{- \frac{\beta}{\gamma} r(t)}, \\
\frac{dr(t)}{dt} &= \gamma \Big [ 1- r(t) - s(0) e^{- \frac{\beta}{\gamma} r(t)} \Big ],
\end{cases} 
\end{align}
to be solved numerically.
At the onset of the epidemic, i.e., for $t\to 0$, the fractions of susceptible and recovered are approximately constant and equal to $s(t) \approx 1$ and $r(t)\approx 0$, respectively. As $t\rightarrow \infty$, if a steady-state regime for $r(t)$ emerges, we expect $\frac{dr(t)}{dt}=0$. Imposing this condition and denoting by 
$r(\infty)=\lim_{t\to\infty} r(t)$ and $s(\infty)=\lim_{t\to\infty} s(t)$, from the second of~\eqref{eq:SIRsol}, we have $r(\infty)=1 - s(0) e^{- \frac{\beta}{\gamma} r(\infty)} \approx 1 - e^{- \frac{\beta}{\gamma} r(\infty)}$, which can be numerically solved for $r(\infty)$. An analytical solution is also available, in terms of the so-called Lambert $W$ function (also known as product logarithm):\cite{Wang2010}
\begin{align}
r(\infty)= \frac{W \left(-\frac{\beta}{\gamma} e^{-\frac{\beta}{\gamma} }\right) + \frac{\beta}{\gamma}}{\frac{\beta}{\gamma}}.
\end{align} 
From the first of~\eqref{eq:SIRsol}, we see that the fractions of recovered and susceptible individuals reach steady-state constant values $r(\infty)$ (typically $>0$) and $s(\infty)$ ($<1$) verifying $r(\infty)=1-s(\infty)$, while $i(\infty)=\lim_{t\to\infty} i(t)=0$.
Suppose $s(t) \approx s(t^\star)$, for some~$t^\star$. For instance, this happens at the beginning of the epidemic with $s(t^\star) =s(0) \approx 1$, and at the end of the epidemic, with  $s(t^\star)=s(\infty)<1$. 
Thus, at different stages of the epidemic, $s(t)$ can be approximately assumed constant over short intervals of time. We make the assumption that variations of $s(t)$ can indeed be neglected over short intervals of time. 
With $s(t)=s(t^\star)$, from the second equation in~\eqref{eq:SIR}, we get
\begin{align} \label{eq:expo}
&\frac{di(t)}{dt} = \alpha \, i(t) \quad \Rightarrow \quad i(t)=i(0) e^{-\alpha t}, 
\end{align}
with $\alpha \dfz \beta s(t^\star)-\gamma$.
Usually, $\beta-\gamma>0$, while $\beta s(\infty)-\gamma<0$, so that $i(t)$ initially grows exponentially at rate $\beta-\gamma$ and eventually decreases exponentially to zero at rate
$\beta s(\infty)-\gamma$. 

\subsubsection*{Proposed Model} \label{sec:discreteSIR}

Epidemic data are typically collected on a daily basis, and a discrete version of~\eqref{eq:expo} with discretization step $\Delta t= 1$ day can be obtained as follows. Let $n$ be the day index. With the obvious notation $i_n=i(n \Delta t)$ and similarly for other quantities, the differential equation on the left-hand side of~\eqref{eq:expo} is approximated by $\frac{i_{n+1} -i_{n}}{\Delta t}=i_{n+1} -i_{n}= \alpha \, i_{n}$. This yields
\begin{align} \label{eq:discretei}
\Delta i_n \dfz i_{n+1}- i_{n} = \alpha \, i_{n} \quad \Rightarrow \quad i_n= i_0 (1+\alpha)^n, \quad n\ge 1,
\end{align}
for some $i_0>0$, and therefore the ratio (also known as growth, for $\alpha > 0$, or decay, for $\alpha < 0$, rate)
\begin{align} \label{eq:x}
x_n \dfz \frac{i_{n}}{i_{n-1}}, \quad n \ge 1,
\end{align}
is constant and equal to $(1+\alpha)$. It is useful to bear in mind that we typically have $|\alpha| \ll1$. From~\eqref{eq:discretei} we see that the sequence~$i_n$ grows or decays exponentially according to the sign of~$\alpha$, and its evolution can be locally assumed to be linear in~$n$ because $(1+\alpha)^n \approx 1+n\, \alpha$, in view of $|\alpha| \ll1$.
For later use, note that, from the third equation in~\eqref{eq:SIR}, we also get
\begin{align} \label{eq:discreter}
\Delta r_n \dfz r_{n+1}- r_{n} = \gamma \, i_{n}.
\end{align}

It is evident that real-world data lead to ``noisy'' versions of the previous expressions. 
There exist two main sources of uncertainty. The first is implicit in the nature of the contagion, which obviously depends on a number of medical and social factors that are problematic to model in deterministic terms. 
Accordingly, the sequence $\{x_n\}_{n \ge 1}$ in~\eqref{eq:x} is better represented by a collection of random variables, leading to the following model for the fraction of infected individuals on day $n$:
\begin{align} \label{eq:modbas}
i_{n}=i_{n-1} \, x_{n} = i_0 \prod_{k=1}^{n} x_k, \qquad n \ge 1.
\end{align}
In~\eqref{eq:modbas},  we assume that $\{x_n\}_{n\ge 1}$ is a sequence of nonnegative \emph{independent} random variables with mean close to unity due to the fact that $|\alpha| \ll 1$.
The independence assumption is crucial to ensure mathematical tractability; however, as we show in the section on ``Performance Assessment - Synthetic Data,'' slight deviations from the condition of perfect independence do not
significantly affect the performance of the proposed Mean-Agnostic Sequential Test (MAST).

The second source of uncertainty is related to the way in which the epidemic data are collected, as we describe next.
First, let us consider the values of  ``new positives'' recorded on day $n$ and its relationship to the ``total cases'' on days $n$ and $n-1$:
\begin{align} \label{eq:shouldbe}
\textnormal{$p_n \dfz$ new positives$_n$ = total cases$_{n}$ - total cases$_{n-1}$}.
\end{align}
Since the number of total cases on day $n$ is equal to $i_n+r_n$, we have $p_n=i_n+r_n-(i_{n-1}+r_{n-1})=\Delta i_{n-1} + \Delta r_{n-1}=(\alpha+\gamma) i_{n-1}$, because of~\eqref{eq:discretei} and~\eqref{eq:discreter}. Then, we get
\begin{align} \label{eq:datacollect}
\frac{p_{n+1}}{p_{n}}=\frac{i_{n}}{i_{n-1}}=x_n,
\end{align}
yielding 
\begin{align} \label{eq:modbasp}
p_{n+1}=p_{n} \, x_{n} = p_1 \prod_{k=1}^{n} x_k, \qquad n \ge 1,
\end{align} 
for some initial value $p_1$. This is important, as while there is a natural modeling in terms of the proportion of infected individuals $i_n$, the time-sequence $\{i_n\}$ is difficult to obtain as the number that get removed from it (by recovery or death) is often not reported. On the other hand, the number of newly infected $p_n$ is easily available, and reported with vigor. 
Based on~\eqref{eq:modbasp}, the sequence $\{x_n\}$ is obtained from $\{p_n\}$ using the relationship $x_n= p_{n+1}/p_{n}$. Returning to the problem of the second source of uncertainty, the sequence of daily new positives that is 
available from the John Hopkins University~\cite{covid-19-JHU} 
is expected to differ from the \emph{actual} numbers of new positives $\{p_n\}_{n \ge 1}$ due to gross errors, missing values or delays in the reported data. For instance, it may happen that, on a given day, the fraction of new positives is not reported (some peripheral data collection unit did not communicate with the central unit in a timely manner) and such unreported fraction is then added to the value observed on a later day. Also, it typically happens that the number of reported cases over the weekend is systematically smaller than the values recorded on other days of the week, due to the smaller number of swabs tested during weekends. Similar periodic modulation effects are already reported in the literature.~\cite{AdaptiveBayesian} 
For these reasons, we first eliminate from the data anomalous values (negative numbers) and then, in place of~\eqref{eq:modbasp}, we consider a more sensible observation model: 
\begin{align} \label{eq:perturbed}
\widetilde p_{n}= p_{n}  + w_{n},  
\qquad n\ge 1,
\end{align}
where $w_n$ is a noise term taking into account the effect of gross errors.
Due to their nature, we expect that gross errors are washed away by averaging the data over a time window of length $L$, where $L$ is of the order of a week or longer. On the other hand, we have seen that the expected behavior of $p_n$ over a sufficiently small time window is approximately linear. For this reason, the downloaded sequence of new positives is first filtered by a moving average filter of length $L$ days (in the paper we use $L=21$, see below).
Under these assumptions, we conclude from~\eqref{eq:perturbed} that:
\begin{align}
\langle \widetilde p_{n} \rangle_L =\langle p_{n} \rangle_L + \langle w_{n} \rangle_L \approx p_n,
\label{eq:brute}
\end{align}
where $\langle  p_n \rangle_L = \frac 1  L \sum_k  p_k$ denotes the $L$-th order uniformly-weighted moving average [MA($L$), for short] of the sequence $\{p_n\}$, where the sum involves~$L$ entries centered on the $n$-th sample. Admittedly, the approximation in \eqref{eq:brute} is
rather sharp, but works for our purposes. Summarizing, the sequence of daily new positives (after removing negative entries) is first processed by an arithmetic moving average filter of order $L$, yielding $\{p_n\}_{n \ge 1}$ and then, from this, the sequence $\{x_n=p_{n+1}/p_{n}\}$ is obtained as in~\eqref{eq:datacollect}.

\begin{table}[!t]

\centering

\begin{tabular}{ l || c | c }
\hline
\rule[-1.3mm]{0mm}{5.5mm}       \textsc{Country}    &       \textsc{p-Value}    &   $\sigma$    \\[1mm] \hline\hline
\rule[-1.0mm]{0mm}{4.5mm}       Albania                         &   $.070$              &   $.020$      \\[0mm]
\rule[-1.0mm]{0mm}{4.5mm}       Austria                         &   $.424$              &   $.025$      \\[0mm]
\rule[-1.0mm]{0mm}{4.5mm}       Belgium                             &   $.104$              &   $.027$      \\[0mm]
\rule[-1.0mm]{0mm}{4.5mm}       Canada                          &   $.030$              &   $.018$      \\[0mm]
\rule[-1.0mm]{0mm}{4.5mm}       France                          &   $.013$              &   $.065$      \\[0mm]
\rule[-1.0mm]{0mm}{4.5mm}       Germany                         &   $.017$              &   $.023$      \\[0mm]
\rule[-1.0mm]{0mm}{4.5mm}       Hungary                             &   $.063$              &   $.032$      \\[0mm]
\rule[-1.0mm]{0mm}{4.5mm}       Italy                               &   $.105$              &   $.015$      \\[0mm]
\rule[-1.0mm]{0mm}{4.5mm}       Netherlands                         &   $.117$              &   $.016$      \\[0mm]
\rule[-1.0mm]{0mm}{4.5mm}       Norway                              &   $.056$              &   $.033$      \\[0mm]
\rule[-1.0mm]{0mm}{4.5mm}       Portugal                            &   $.623$              &   $.017$      \\[0mm]
\rule[-1.0mm]{0mm}{4.5mm}       Spain                           &   $.001$              &   $.047$      \\[0mm]
\rule[-1.0mm]{0mm}{4.5mm}       UK                              &   $.006$              &   $.016$      \\[0mm]
\rule[-1.0mm]{0mm}{4.5mm}       US                                  &   $.662$              &   $.006$      \\[1mm] \hline
\end{tabular}

\caption{$p$-value of Kolmogorov-Smirnov goodness-of-fit test, and estimated value of $\sigma$.}
\label{tab:ks-test}

\vspace{-3mm}

\end{table}

\subsubsection*{Statistical Characterization}
Recall that the subscript $n$ to $x_n$ and other quantities denotes the index of the day. The statistical distribution of the sequence $\{x_n\}$ is derived 
from COVID-19 data made available by the Johns Hopkins University,~\cite{covid-19-JHU} as described next.
For concreteness, let us refer to the data from Italy. Figure~2(a) of the main document shows the measured sequence (not normalized to  total population) of new positives $\{\widetilde p_n\}$ and in green the smoothed version $\{p_n\}$ obtained by MA(21), i.e., as the output of  a moving average filter of length $L=21$ days, fed by $\{\widetilde p_n\}$. The corresponding sequence of ratios $\{x_n=p_{n+1}/p_{n}\}$ is shown in Fig.~2(b) in green.
Numerical investigations, not detailed for the sake of brevity, show that the growth rate $x_n$ can be modeled by a Gamma random variable with shape parameter $\kappa$ and scale parameter $\theta$, with $\kappa \theta$ close to unity. Since $\kappa$  is on the order of several hundreds, the Gamma distribution is practically indistinguishable from a Gaussian distribution with mean $\mu=\kappa \theta$ on the order of unity, and standard deviation $\sigma=\sqrt{\kappa} \theta \ll 1$. In light of these considerations,
we adopt the Gaussian model because it leads to a detector with simple and intuitive structure.

Accordingly, let us reconsider the sequence $\{x_n\}$ shown in green in Figure~2(b).
First, we disregard the initial portion of the sequence $\{x_n\}$ that corresponds to the first peak of $\{p_n\}$ (first wave of the epidemic). The omitted data corresponds to values of the day index $n$ smaller than the index of the first passage (from the above) by one of the sequence $\{x_n\}$. Then, we consider the smoothed version $\{\widehat \mu_n\}$ of $\{x_n\}$, shown in magenta in Figure~2(b).
The smoothing is obtained by processing $\{x_n\}$ again by a MA(21) filter. The length $L=21$ of the two MA filters used in this analysis have been selected by trial and error. 
Note that the filters are not \emph{causal}:\cite{oppenheim-SaS} to compute the output at time $n$, $(L-1)/2$ successive samples of the input are needed.
Note also that the filters are truncated at the endpoints where less than $L$ samples are available.

\begin{figure}
\centering
\psfragfig[width=0.72\textwidth]{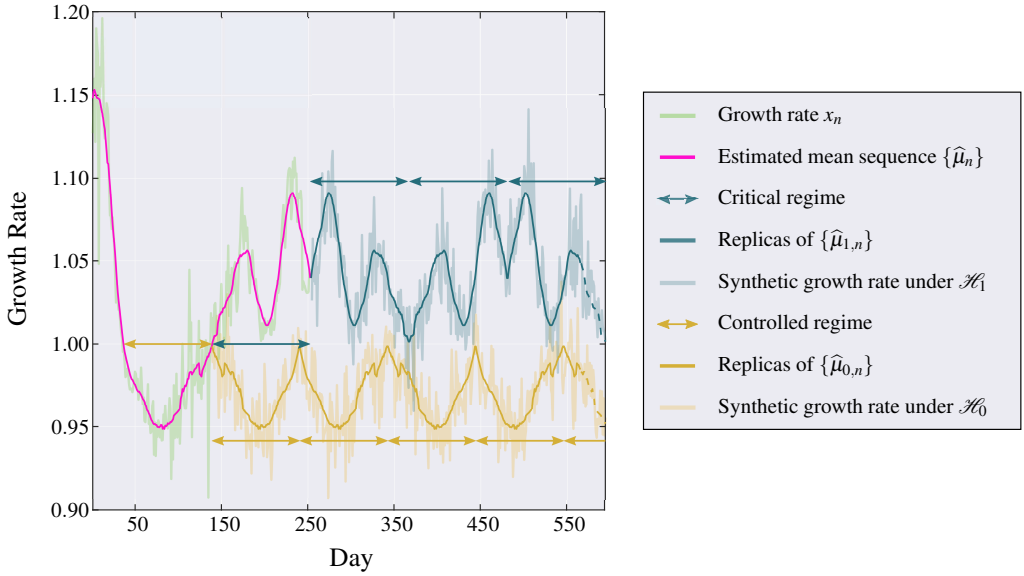}
\vspace{6mm}

\caption{Example of the construction of the periodic counterparts of $\{ \widehat{\mu}_{0,n} \}$ and $\{ \widehat{\mu}_{1,n} \}$ from the Italian data, used to generate the synthetic grow rates under the controlled and the critical regime, respectively.
From the growth rate sequence $x_{n}$ (green curve), the estimated mean sequence $\{ \widehat{\mu}_{n} \}$ (magenta) is obtained through a uniformly-weighted moving average filter of length  21 days. 
From $\{ \widehat{\mu}_{n} \}$, the two sub-sequences $\{\widehat{\mu}_{0,n} \}$ (dark yellow) and $\{\widehat{\mu}_{1,n} \}$ (dark cyan) are extracted, so as to verify  $\widehat{\mu}_{0,n} \leqslant 1$ and $\widehat{\mu}_{1,n} > 1$.
The two sub-sequences are then replicated, with the odd replicas flipped to preserve continuity, yielding the dark yellow and dark cyan solid curves, which are finally used as mean values to generate the synthetic data under the controlled and the critical regime, respectively.
One realization of the resulting synthetic growth rates under~${\cal H}_{0}$ and under~${\cal H}_{1}$ is superimposed to the mean values, for illustration purpose.}
\label{fig:mean-replicate-construction}

\end{figure}

By subtracting the smoothed version $\{\widehat \mu_n\}$ from the sequence $\{x_n\}$, a sequence $\{x_n-\widehat \mu_n\}$ with approximately zero-mean is obtained. 
Then, we verify that the random variables $x_n- \widehat \mu_n$, $n=1,2,\dots$, have one and the same standard deviation $\sigma \ll 1$, with good accuracy.
In particular, there is no evidence of a substantial change in standard deviation between the regions with $\widehat \mu_n \le 1$ and those in which $\widehat \mu_n >1$.
The estimated values of $\sigma$ for the 14 nations considered in the main document is reported in Table~\ref{tab:ks-test}.
To distinguish between these two regions, we introduce the notations $\mu_{0,n} \dfz \mu_n \le 1$ and $\mu_{1,n} \dfz \mu_n>1$ for the actual (unknown) mean values. Note that, since $\mu_{0,n}$ and $\mu_{1,n}$ are close to unity and $\sigma \ll 1$, we see that $x_k < 0$ with negligible probability.

The quantity $\{x_n-\widehat \mu_n\}$ so obtained is modeled as a sequence of zero-mean independent Gaussian random variables with (nation-dependent but small) standard deviation $\sigma$.
The independence assumption is necessary to ensure mathematical tractability and, in practice, slight deviations from the condition of perfect independence do not significantly affect our results.
In this respect, it should be also noted that the smoothing operation shown in Eq.~\eqref{eq:brute} enforces correlation over the sequence $\{p_n\}$, and, in turn, over the sequence of ratios $\{x_n=p_{n+1}/p_n\}$.

Finally, in Table~\ref{tab:ks-test}, we verify that the $p$-value obtained by the Kolmogorov-Smirnov (KS) goodness-of-fit test\cite{feller1948} is larger than $0.01$, meaning that the Gaussian assumption cannot be rejected at $1\%$ significance level, for almost all the countries (except Spain and UK).
We thus arrive at the following model of the observable
growth rate sequence $\{x_n\}$: 
$x_n \sim \N(\mu_{0,n}, \sigma)$ with $\mu_{0,n} \le 1$, under the controlled regime, and $x_n \sim \N(\mu_{1,n}, \sigma)$ with $\mu_{1,n}>1$, under the critical one.

\subsection*{Performance Assessment}

\subsubsection*{Real Data}

The implementation of the MAST algorithm does not require knowledge of the mean sequences $\{\mu_{0,n}\}$ and $\{\mu_{1,n}\}$ and, in fact, it has been designed just to cope with this uncertainty. However, the performance of the test depends, of course, on the specific scenario under consideration. 
Performance of
MAST run on publicly available COVID-19 infection data from different countries
is reported in the main document and obtained as follows.
For each country, we investigate the performance of
MAST by using the estimates $\{ \widehat{\mu}_{0,n} \}$ and $\{ \widehat{\mu}_{1,n} \}$  of the mean sequences $\{\mu_{0,n}\}$ and $\{\mu_{1,n}\}$. Since we need to simulate arbitrarily long data streams under both the controlled and critical regimes, we construct periodic counterparts of the sequences $\{ \widehat{\mu}_{0,n} \}$ and $\{ \widehat{\mu}_{1,n} \}$, as illustrated in Fig.~\ref{fig:mean-replicate-construction}.
Using these periodic counterparts, we
obtain the performance of the test by standard Monte Carlo computer experiments,~\cite{kaydetection, Allen2017, Lehmann-testing} involving $10^5$ independent runs for each value of the threshold $\chi$. This
gives the relationship between $R$ and $\chi$, and between $\Delta$ and $\chi$, for relatively small values of $\chi$. The former relationship is almost exactly exponential, and the latter almost exactly linear. Thus, the aforementioned relationships are  extrapolated to arbitrarily large values of $\chi$. This explains why the figures of the main document report values of $R$ that would be difficult to obtain by standard computer experiments.

\subsubsection*{Synthetic Data}

\begin{figure}

\centering

\begin{minipage}{0.45\textwidth}
    \centering
    \hspace{-10mm}\psfragfig[width=.85\textwidth]{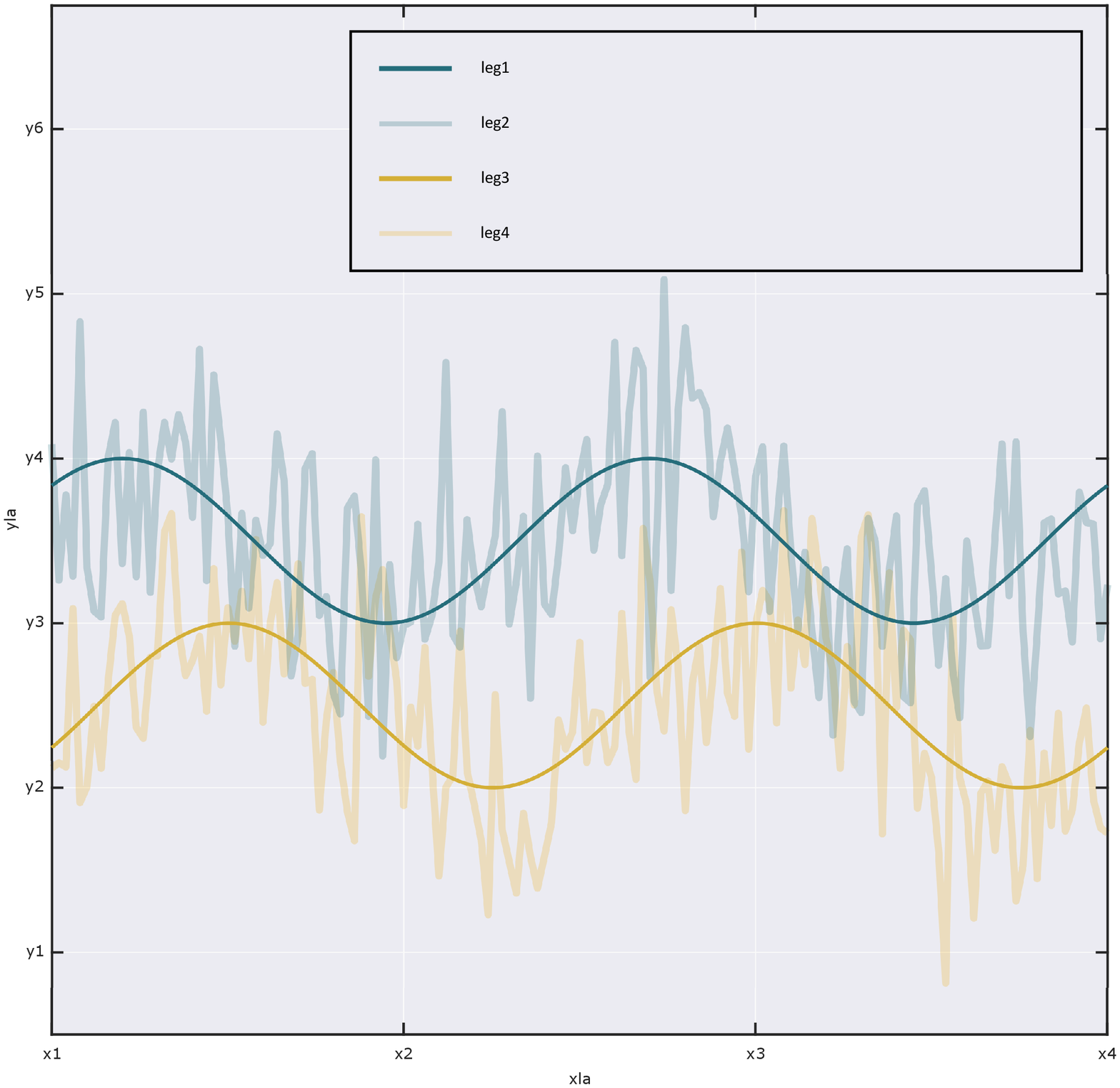} \\[5.5mm] \hspace{-1.5mm} (a)
\end{minipage}%
\begin{minipage}{0.45\textwidth}
    \centering
    \hspace{10mm}\psfragfig[width=0.85\textwidth]{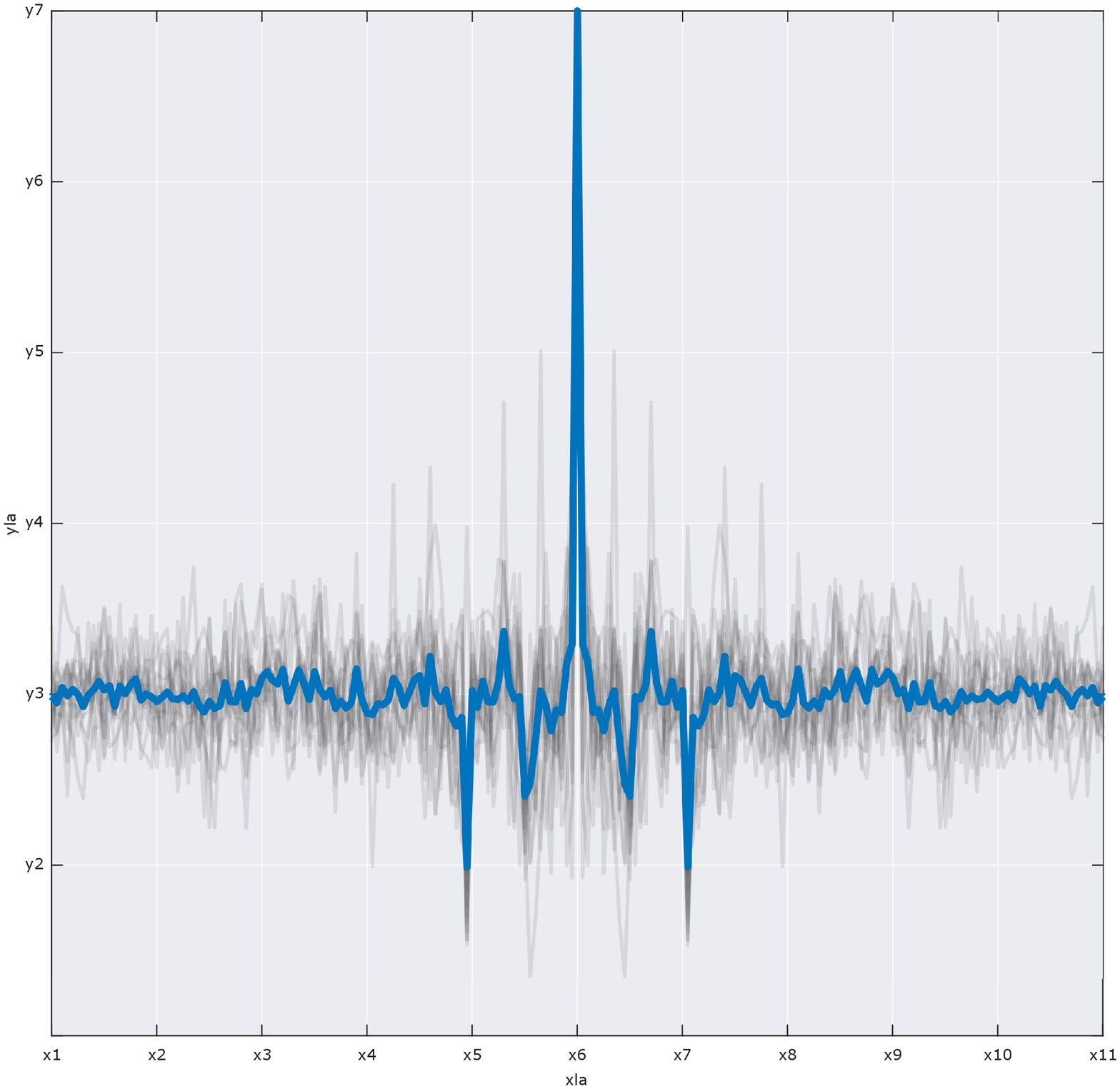} \\[5.5mm] \hspace{1.5mm} (b)
\end{minipage}%

\vspace{4mm}

\caption{(a) Mean sequences $\{ \mu_{0,n} \}$ (dark yellow) and $\{ \mu_{1,n} \}$ (dark cyan) used in the synthetic scenario.
These are sinusoidal waves with period of 75 days
oscillating between $1 - \varepsilon$ and $1$ under ${\cal H}_{0}$, and between $1$ and $1 + \varepsilon$ under ${\cal H}_{1}$, with $\varepsilon = 0.1$.
One realization of the resulting synthetic growth rates under ${\cal H}_{0}$ and under ${\cal H}_{1}$ is superimposed to the mean values, for illustration purpose.
(b) Normalized correlation (blue line) enforced
on the synthetic sequence $\{ x_n - \mu_{n} \}$ used to numerically evaluate the performance of
MAST.
This normalized correlation is derived by averaging the normalized correlations obtained from the data of the 14 nations considered in the main document; each of these is reported on the background (light grey lines).}

\label{fig:sim-means}

\end{figure}

\begin{figure}

\centering
\psfragfig[width=.9\textwidth]{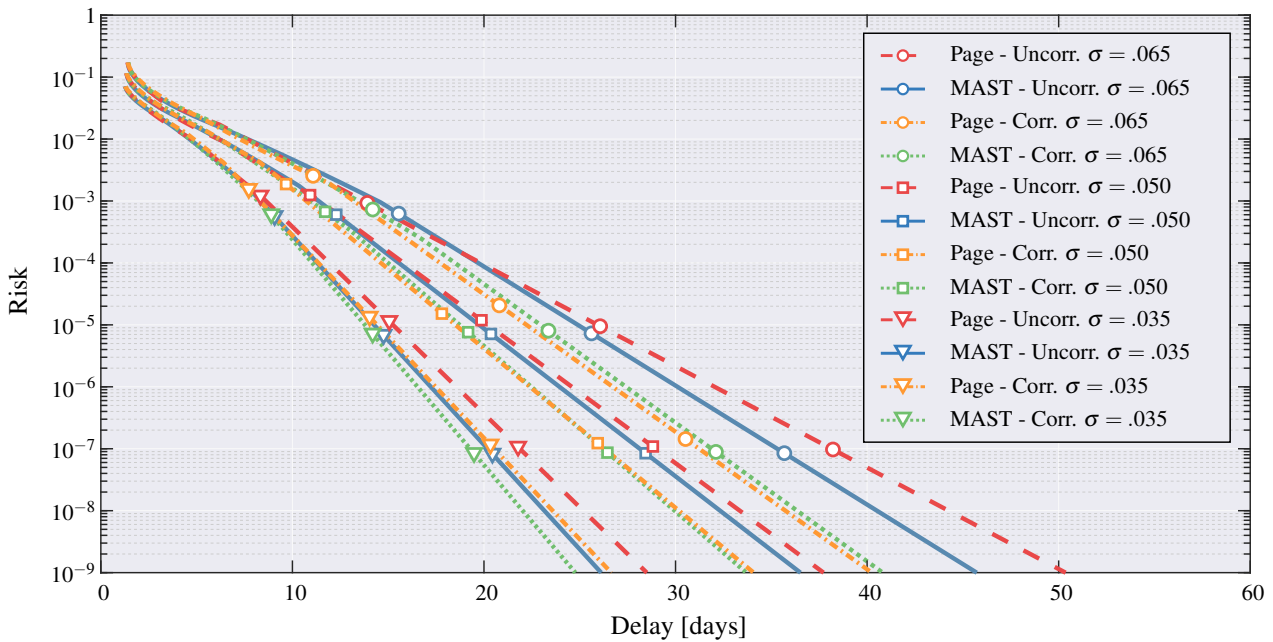}

\vspace{4mm}

\caption{Performance of
MAST and the naive Page's test in terms of risk versus mean delay.
Red dashed lines and blue solid lines refer to the naive Page's test and
MAST, respectively, when run on uncorrelated sequences;
yellow dash-dotted lines and green dotted lines refer to the naive Paige's test and
MAST, respectively, when applied on correlated sequences.
The markers allow to distinguish between different values of $\sigma$: circles for $\sigma = 0.065$, squares for $\sigma = 0.050$, and triangles for $\sigma = 0.035$.}

\label{fig:sim-performance}

\end{figure}

The performance of
MAST
is also evaluated in a synthetic scenario and compared with those of a \textit{naive} Page's test.
To emulate a realistic scenario, we assume that the mean sequence $\{ \mu_{n} \}$ has a periodic evolution not exceeding $1$ in the controlled regime, and not falling below $1$ in the critical regime.
Specifically, we set
\begin{align}
    \mu_{0,n} &= 1 + \frac{\varepsilon}{2} \big( \mathrm{cos}(2\pi n M^{-1} + \phi_{0}) - 1 \big) \, ,
    \intertext{and}
    \mu_{1,n} &= 1 + \frac{\varepsilon}{2} \big( \mathrm{cos}(2\pi n M^{-1} + \phi_{1}) + 1 \big) \, .
\end{align}
Fig.~\ref{fig:sim-means}(a) shows the mean sequences used for this analysis, with $\varepsilon = 0.1$, period $M = 75$ days, and phases $\phi_{0}$ and $\phi_{1}$ uniformly drawn
from $[0,2\pi)$; in addition, one realization of the resulting synthetic growth rates under ${\cal H}_{0}$ and under ${\cal H}_{1}$ is provided, obtained by adding to each mean sequence a zero-mean Gaussian noise process with standard deviation $\sigma = 0.050$.
Moreover, in order to numerically demonstrate that slight deviations from the condition of perfect independence of the sequence $\{ x_n - \mu_{n} \}$ do not significantly affect the performance of the MAST (as
claimed above), we also consider the case of a correlated zero-mean Gaussian noise process.
The (normalized) correlation we enforce
on the sequence $\{ x_n - \mu_{n} \}$ is shown in blue in Fig.~\ref{fig:sim-means}(b); this is obtained by averaging the
correlations computed from the sequences $\{ x_n - \widehat{\mu}_{n} \}$ of the 14 nations considered in the main document.
The correlated Gaussian
process is generated by filtering an uncorrelated Gaussian
sequence as described in~\cite[Ch. 11.3.2]{OppVer:B16}.

MAST is compared with a naive Page's test that is aware of the bounds of the mean sequence under each hypothesis, namely, the lower bound $1 - \varepsilon$ and the upper bound $1$ under ${\cal H}_{0}$, and the lower bound $1$ and the upper bound $1 + \varepsilon$ under ${\cal H}_{1}$;
naive here implies that the test does not know the exact evolution of the mean sequence, but only its extreme values.
The naive Page's test is based on Eq. (6), reported in the main document, imposing $\alpha = \epsilon$, and
corresponds to the standard Page's test when $M = 1$, $\phi_{0} = \pi$, $\phi_{1} = 0$, and the data are uncorrelated.

Fig.~\ref{fig:sim-performance} presents the performance of
MAST and the naive Page's test in terms of risk versus mean delay for different values of $\sigma$, obtained
by standard Monte Carlo computer experiments~\cite{kaydetection, Allen2017, Lehmann-testing} involving $10^5$ independent runs for each value of the threshold $\chi$.
We observe that in the case of uncorrelated sequences,
MAST outperforms the naive Page's test, even though the former does not have any prior information about the mean sequences, except for them being above or below $1$, that is, $\mu_{0,n} \leqslant 1$ and $\mu_{1,n} > 1$.
By enforcing the correlation, the performance of both
MAST and the naive Page's test improves; however, there are no remarkable differences, with the displacement in terms of mean delay being no more than 1 day for the MAST and 2 days for the naive Page's test at a risk level of $10^{-4}$.

\end{document}